\documentclass[traditabstract,twocolumns]{aa}
\usepackage{graphicx}	
\usepackage{amsmath}	
\usepackage{amssymb}	
\usepackage{xcolor}     
\usepackage{soul}       
\usepackage{array}
\usepackage{txfonts}
\usepackage[colorlinks=true,linkcolor=blue,citecolor=blue, urlcolor=blue]{hyperref}
\bibpunct{(}{)}{;}{a}{}{,}

\usepackage{fontawesome}  
\usepackage{lipsum}
\usepackage{multirow}
\usepackage{colortbl}  

\usepackage{rotating}
\usepackage{adjustbox}

\definecolor{fredcolor}{RGB}{204, 204, 0}
\definecolor{changesgreyblue}{RGB}{78, 129, 166}

\newcommand{\herculens}{\textsc{Herculens}\xspace}
\newcommand{\lenstro}{\textsc{Lenstronomy}\xspace}
\newcommand{\molet}{\textsc{MOLET}\xspace}
\newcommand{\jax}{\textsc{JAX}\xspace}

\newcommand{\Hc}{\ensuremath{H_0}\xspace}
\newcommand{\op}[1]{\ensuremath{\boldsymbol{\mathsf{#1}}}\xspace}
\newcommand{\mat}[1]{\ensuremath{{\rm \bf #1}}\xspace}
\newcommand{\bfsym}[1]{\ensuremath{{\boldsymbol #1}}\xspace}
\newcommand{\source}{\bfsym{s}}
\newcommand{\lenslight}{\bfsym{\ell}}
\newcommand{\sourcescalar}{\ensuremath{s}\xspace}
\newcommand{\lenslightscalar}{\ensuremath{\ell}\xspace}

\newcommand{\lenspot}{\bfsym{\psi}}
\newcommand{\lenspotscalar}{\ensuremath{\psi}\xspace}

\newcommand{\lenspotpix}{\bfsym{\psi_{\rm pix}}}
\newcommand{\convpix}{\bfsym{\kappa_{\rm pix}}}
\newcommand{\sourcepix}{\bfsym{s_{\rm pix}}}
\newcommand{\lenslightpix}{\bfsym{\ell_{\rm pix}}}
\newcommand{\modelparam}{\bfsym{\eta}}
\newcommand{\modelparamsgl}{\ensuremath{\eta}\xspace}
\newcommand{\modelparamMAP}{\ensuremath{\modelparam_{\rm MAP}}\xspace}
\newcommand{\data}{\bfsym{d}}
\newcommand{\model}{\bfsym{m}}
\newcommand{\lensingop}{\ensuremath{\boldsymbol{\mathsf{L}}_{\lenspot}}\xspace}
\newcommand{\lensingopsmooth}{\ensuremath{\boldsymbol{\mathsf{L}}_{\tilde\lenspot}}\xspace}
\newcommand{\convop}{\op{B}}
\newcommand{\noise}{\bfsym{n}}
\newcommand{\covmatrix}{\mat{C}}
\newcommand{\datacovmatrix}{\ensuremath{\covmatrix_\data}\xspace}

\newcommand{\waveletop}{\ensuremath{\bfsym{\Phi}}\xspace}
\newcommand{\starletop}{\ensuremath{\bfsym{\Phi}_{\rm st}}\xspace}
\newcommand{\battlelemarieop}{\ensuremath{\bfsym{\Phi}_{\rm bl}}\xspace}

\newcommand{\weights}{\ensuremath{\mat{W}}\xspace}
\newcommand{\zd}{\ensuremath{z_{\rm d}}\xspace}
\newcommand{\zs}{\ensuremath{z_{\rm s}}\xspace}
\newcommand{\proba}[1]{\ensuremath{P\left(#1\right)}\xspace}
\newcommand{\prior}[1]{\ensuremath{\mathcal{P}\left(#1\right)}\xspace}
\newcommand{\priorspec}[2]{\ensuremath{\mathcal{P}_{#1}\left(#2\right)}\xspace}
\newcommand{\loss}{\ensuremath{L}\xspace}
\newcommand{\gradloss}{\ensuremath{\nabla\loss}\xspace}
\newcommand{\hessianloss}{\ensuremath{\mat{H}_\loss}\xspace}
\newcommand{\expfim}{\ensuremath{\mathcal{I}}\xspace}
\newcommand{\obsfim}{\ensuremath{\mat{I}}\xspace}
\newcommand{\likelihood}[1]{\ensuremath{\mathcal{L}\left(#1\right)}\xspace}

\newcommand{\priorsimple}{\ensuremath{\mathcal{P}}\xspace}
\newcommand{\likelihoodsimple}{\ensuremath{\mathcal{L}}\xspace}

\newcommand{\normone}[1]{\ensuremath{\left\lVert \,#1\, \right\rVert_1}}

\DeclareMathOperator*{\argmin}{arg\,min}

\begin{document}


\title{Using wavelets to capture deviations from smoothness\\in galaxy-scale strong lenses}

\titlerunning{Capturing deviations to smoothness with wavelets}

\author{
A.~Galan\inst{\ref{epfl}}\fnmsep\thanks{aymeric.galan@epfl.ch}, G.~Vernardos\inst{\ref{epfl}}, A.~Peel\inst{\ref{epfl}}, F. Courbin\inst{\ref{epfl}}, J.-L. Starck\inst{\ref{cea}}}

\institute{
Institute of Physics, Laboratory of Astrophysics, Ecole Polytechnique 
F\'ed\'erale de Lausanne (EPFL), Observatoire de Sauverny, 1290 Versoix, 
Switzerland \label{epfl}
\goodbreak
\and
AIM, CEA, CNRS, Universit\'e Paris-Saclay, Universit\'e de Paris, F-91191 Gif-sur-Yvette, France \label{cea}
}

\abstract{
Modeling the mass distribution of galaxy-scale strong gravitational lenses is a task of increasing difficulty. The high-resolution and depth of imaging data now available render simple analytical forms ineffective at capturing lens structures spanning a large range in spatial scale, mass scale, and morphology. In this work, we address the problem with a novel multiscale method based on wavelets. We tested our method on simulated Hubble Space Telescope (HST) imaging data of strong lenses containing the following different types of mass substructures making them deviate from smooth models: (1) a localized small dark matter subhalo, (2) a Gaussian random field (GRF) that mimics a nonlocalized population of subhalos along the line of sight, and (3) galaxy-scale multipoles that break elliptical symmetry. We show that wavelets are able to recover all of these structures accurately. This is made technically possible by using gradient-informed optimization based on automatic differentiation over thousands of parameters, which also allow us to sample the posterior distributions of all model parameters simultaneously. By construction, our method merges the two main modeling paradigms---analytical and pixelated---with machine-learning optimization techniques into a single modular framework. It is also well-suited for the fast modeling of large samples of lenses. All methods presented here are publicly available in our new \herculens package \href{https://github.com/austinpeel/herculens}{\faGithub}.
}


\maketitle

\section{Introduction \label{sec:intro}}

The Lambda cold dark matter cosmological model encapsulates our current understanding of the Universe, accurately explaining a number of observations on large scales ($>\!10$ Mpc), such as the cosmic microwave background temperature and polarization anisotropies \citep{Planck2020}, baryon acoustic oscillations \citep[e.g.,][]{Raichoor2021}, and the accelerated expansion of the Universe \citep[e.g.,][]{Riess2020}. The dark matter (DM) component of this model plays a major role in the hierarchical collapse of matter due to gravitational instability, which eventually produced the galaxies populated by stars that we observe today \citep[e.g.,][]{Toomre1972,Dubinski1994,Springel2006}.
Despite its successes on large scales, the effect of DM on galactic and subgalactic scales ($<10$ Mpc), where its interplay with baryons is mediated by nonlinear and poorly understood mechanisms \citep[e.g.,][]{Blumenthal1986,ZubovasKing2012}, poses a challenge in explaining observations \citep[e.g.,][]{Vogelsberger2020}. For instance, predictions from the cold DM model lead to overly cuspy central density profiles \citep[the so-called cusp-core problem,][]{Moore1994,deBlock2010}, too many low-mass ($\lesssim\!10^8\ {\rm M}_\odot$) dwarf galaxies \citep[the missing satellites problem,][]{Moore1999,Klypin1999} and intermediate-mass ($\sim\!10^{10}\ {\rm M}_\odot$) galaxies populated by too few stars \citep[the too-big-to-fail problem,][]{BoylanKolchin2011,Papastergis2015}.

One path to reconcile theory with observations is to examine alternative DM models that assume different properties of the DM particle. For example, warm DM simulations \citep[e.g.,][]{Li2017} have been able to produce far fewer dwarf galaxy satellites, which in line with observations. Another path is to improve our ability to detect observational signatures of DM through its gravitational influence on baryonic matter, such as those observed in the local Universe within the Milky Way \citep[e.g., with stellar streams,][]{Erkal2016} and its neighborhood \citep[e.g., within dwarf galaxy satellites,][]{Nadler2021}. At cosmological distances, gravitational lensing is a direct and more efficient probe of the mass in galaxies, and has the potential to measure DM effects down to kpc in size and $10^7\ {\rm M}_{\odot}$ in mass \citep[e.g.,][]{Hezaveh2016b,ChatterjeeKoopmans2018,Gilman2019}. This is a crucial range of mass where most DM models tend to disagree in their predictions.

Galaxy-galaxy strong lensing occurs when two galaxies at different redshifts align along the line of sight, with the foreground galaxy (the lens) deflecting incoming light rays of the background one (the source). From the observer’s perspective, the source appears magnified, distorted, and split into multiple images. By modeling the observed lensed features through the well-established physics of the lens equation, one can constrain the total mass distribution in galaxies and measure their DM content and formation history.
However, lens modeling is an under-constrained problem, mostly because both the lens mass and the source light distributions are a priori unknown, and the lensed source is often blended with the lens light. Assumptions are therefore needed to reconstruct the lensing mass, the lens light, and the source light distributions by inverting the lens equation. These assumptions are often priors on the shape of mass and light profiles, or on their higher order statistical properties. Currently used analytical profiles are sufficient to capture first-order properties of the lens mass distribution (e.g., power-law profiles), but they lack the degrees of freedom to capture those small-scale features that are critical for determining DM properties \citep[e.g.,][]{He2022}.

Increasing the complextiy of a lens model is not trivial due to degeneracies between the lens potential and the source light: a complex structure in the observed lensed features could be attributed to either local perturbations in the potential or to an intrinsically complex source light distribution (e.g., clumpy star-forming galaxies or galaxy mergers).
\citet{Koopmans2005} used a Taylor expansion to address this, deriving a perturbative correction for the lens equation that combines spatial derivatives of both the source and the potential. This approach extends the semilinear inversion technique of \citet{WarrenDye2003} by adding a perturbing field to the smooth lens potential, which does not assume any specific shape and can be solved for in the same way as the source. The lens potential and the source light were discretized and cast on two grids of pixels in the lens and source planes respectively, on which the two fields were reconstructed \citep[see also][for extensions of the technique]{Vegetti2009,Suyu2009,Vernardos2022}.
For the source, various priors have been explored, including analytical functions ranging from the Sérsic profile \citep{Sersic1963} to shapelet basis sets \citep{TagoreKeeton2014,Birrer2015}, Gaussian processes \citep{Karchev2022}, or deep generative models \citep{Chianese2020}. More complex sources can be modeled using grids of pixels, combined with curvature-based \citep{Suyu2006,NightingaleDye2015}, Gaussian process \citep{Vernardos2022}, or wavelet-based \citep{Joseph2019,Galan2021} regularizations.
Recently, \citet{Vernardos2022} studied the effect of specific prior assumptions on both the potential perturbations and the source light distribution, confirming their degeneracy and that the particular choice of priors plays an important role in recovering the underlying potential perturbations.

In a real-world application, we do not know a priori the dominating type of perturbations, such that we need a method that is both flexible yet robust in recovering their main properties. These perturbations may be due to isolated subhalos, or subhalo populations along the line of sight. Both cases allow us to probe the low-mass end of the dark matter mass function, and possibly the shape of DM halos, for which different DM model predictions disagree. The gravitational imaging technique has been used for the detection of subhalos based on the reconstructed potential perturbations. If the pixelated reconstruction contains a well-localized mass over-density, it is replaced by an analytical profile such as a Navarro-Frenk-White (NFW) halo \citep{Navarro1996} whose parameters are further optimized \citep{Vegetti2009}. Using this method, \citet{Vegetti2012} reported the detection of a $10^{8.28}\ {\rm M}_\odot$ dark halo, and \citet{Vegetti2014} used nondetections in eleven systems of the Sloan Lens ACS Survey (SLACS) sample \citep{Bolton2006} observed with the Hubble Space Telescope (HST) to constrain the mean projected substructure mass fraction in the context of cold DM. Using a different method based on comparing model likelihoods between a model that explicitly includes a subhalo at a given position with a model that does not, \citet{Hezaveh2016a} reported the detection of a substructure of mass $10^{8.96}\ {\rm M}_\odot$ from ALMA observations. In general, constraints from individual subhalo detections can mainly be improved with the analysis of larger samples of lenses \citep{Vegetti2014}, and depend on the shape of the profile used to describe the subhalo mass distribution \citep{Despali2022}.

These direct detection methods quickly become too computationally expensive for detecting $\gtrsim2$ subhalos. Considering populations of subhalos instead allows one to probe even lower subhalo masses ($\lesssim10^{7}\ {\rm M}_\odot$) through their collective effects, since low-mass subhalos are predicted to be more numerous \citep{Hezaveh2016b}. A population of subhalos is described in a statistical way, but its properties are directly related to those of DM models, which can thus be constrained \citep[e.g., the thermal relic mass of the warm DM particle,][]{Birrer2017}. One such statistical description of the perturbed lensing potential has been introduced in \citet{ChatterjeeKoopmans2018}, using a power-spectrum analysis of the (lensed) source surface brightness fluctuations, reaching a sensitivity down to a few kpc in the subhalo mass power-spectrum \citep[see also][]{Bayer2018}.
Similarly, \citet{CyrRacine2019} decomposed model residuals into Fourier modes and linked them to the substructure power-spectrum. More recently, \citet{Vernardos2022} used the gravitational imaging technique to reconstruct the perturbing field of a population of subhalos and recovered its power-spectrum properties, especially the slope, remarkably well.
Several studies have also employed deep learning techniques to infer the presence of subhalo populations \citep[e.g.,][]{Brehmer2019,DiazRivero2020,Varma2020,Coogan2020,Vernardos2020,Ostdiek2022,Adam2022}. While it is still unclear if these methods are strongly limited by the simplifying assumptions on their training data (e.g. the absence of lens light, fixed instrumental properties, etc.), they are a promising path forward to speed up computations for application to large samples of lenses, possibly in combination with more classical approaches.

Independently of the presence of subhalos, deviations from smoothness can occur within the lens galaxy mass distribution itself, manifesting as additional radial or azimuthal structures. For instance, deviations along the radial direction can be mass-to-light radial gradients \citep{OldhamAuger2018,Sonnenfeld2018b,Shajib2021}, while angular structures can be the consequence of ellipticity gradients and twists \citep{Keeton2000,VandeVyvere2022}. These cannot be captured by the typically employed elliptical power-law mass models. Free-form techniques have been one way to address these limitations by dismissing the smooth component of the potential and relying solely on a grid of mass pixels governed by a set of priors (either physically or mathematically motivated) to prevent the appearance of un-physical mass distributions \citep{Saha1997,Coles2014}. However, retaining the smooth component already provides a reliable first-order model, from which to explore higher order deviations. Azimuthal structures can be described by higher-order multipoles that have been identified in stellar populations from both real observations and cosmological simulations \citep{Trotter2000,Claeskens2006}. These can be explained by AGN feedback that suppresses the formation of disks in massive galaxies, as they tend to evolve from disky to elliptical or even boxy shapes \citep[i.e., quadrupoles,][]{Frigo2019}. Recently, \citet{VandeVyvere2022_TDC7} demonstrated that quadrupole moments of low amplitude, based on results of \citet{Hao2006}, can be detected in current HST observations of strong lenses, although their detectability depends on numerous factors such as their alignment with the smooth potential or the degrees of freedom in the source model. Using very long baseline interferometric observations of a strong lens, \citet{Powell2022} recently reported the detection of multipole structures beyond ellipticity in the deflector mass distribution. While not accounting for those multipoles in the model can bias the measurement of the lens mass profile up to a few percent \citep{VandeVyvere2022_TDC7,Powell2022}, their detection most importantly holds valuable information on the formation history of galaxies, as signatures of their past evolution.

Our goal in this paper is to unify the modeling of generic lens potential perturbations in a robust framework that includes, but is not limited to, the three categories presented above: individual or populations of subhalos and higher order moments in the lens mass distribution.
To achieve this, we extend our previous work in \citet{Galan2021} by including a reconstruction of lens potential perturbations regularized with wavelets, and demonstrate that our method can successfully reconstruct perturbed lens potentials of different origin.
Our technique benefits from the multiscale properties of the wavelet transform, along with well-motivated sparsity constraints to reconstruct the various spatial scales over which perturbations to the smooth potential can occur.
We implement our method using a fully differentiable algorithm based on automatic differentiation, first introduced by \citet{Wengert1964}.
This programming framework gives direct access to all the derivatives of the highly nonlinear loss function at a negligible computational cost, which enables the use of robust gradient-informed algorithms to explore the parameter space and optimize the model parameters.
As a result, convergence to the maximum-a-posteriori (MAP) solution is fast and first-order error estimates are obtained through Fisher matrix analysis. An efficient exploration of the parameter space for estimating posterior distributions is then performed via gradient-informed Hamiltonian Monte-Carlo (HMC) sampling.

We present our methodology in Sect.~\ref{sec:method}. The simulated examples used to demonstrate the capabilities of our method are presented in Sect.~\ref{sec:experimental_setup}. We perform the reconstruction of perturbations to the lens potential by uniformly modeling these examples and present our results in Sect.~\ref{sec:modeling_results}. We then evaluate the reconstructions  of the perturbations in Sect.~\ref{sec:model_properties}. We conclude this work and discuss its future prospects in Sect.~\ref{sec:discussion_conclusion}.

\section{Methodology \label{sec:method}}

\begin{figure*}
    \centering
    \includegraphics[width=\linewidth]{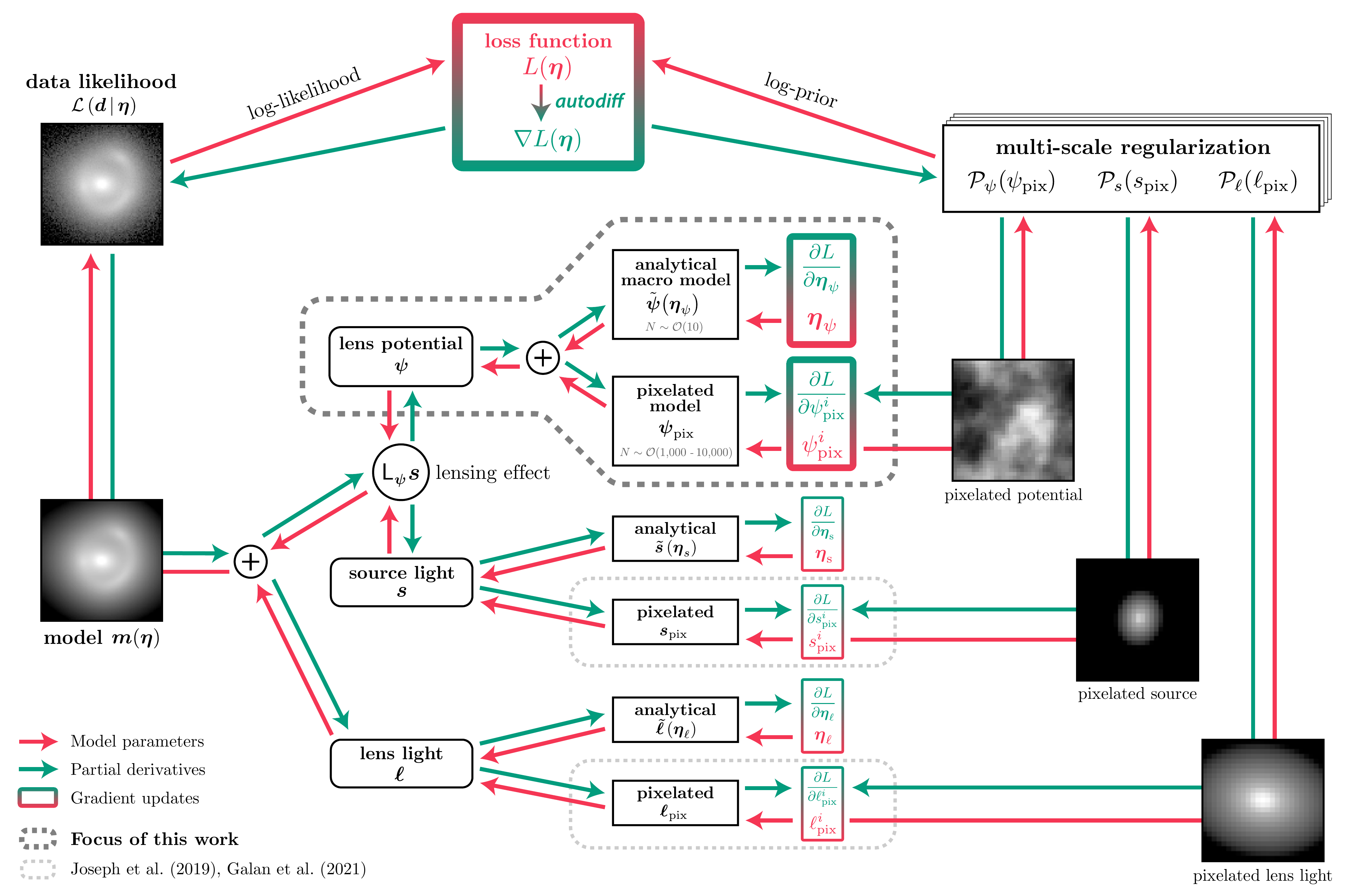}
    \caption{Flow chart of the fully differentiable method used in this work and implemented in \herculens, which merges analytical profiles with pixelated components. We indicate the typical number of parameters $N$ for the analytical potential $\modelparam_\lenspotscalar$ and the pixelated potential \lenspotpix. Partial derivatives are computed and propagated for all parameters. These derivatives are then used to update parameter values in the direction indicated by the gradient of the loss function \loss. We note that the blurring operator \convop (see Eq.~\ref{eq:linear_lens}) is also used to generate the model \model but is not shown in the diagram to avoid clutter. This work focuses on modeling the lensing potential \lenspot, and we refer to \citet{Joseph2019} and \citet{Galan2021} for the modeling of surface brightness distributions.}
    \label{fig:scheme_autodiff_flow}
\end{figure*}

In this section we introduce the gravitational lensing formalism and describe in detail the various aspects of our method. Fig.~\ref{fig:scheme_autodiff_flow} summarizes the main components of the method and can be referred to throughout this section. Some important mathematical notations are also summarized Table~\ref{tab:notations}.

\begin{table}[!h]
\caption{List of the main notations used in the paper.} \label{tab:notations}
\renewcommand{\arraystretch}{1.4}
\centering
{\small
\begin{tabular}{ll}

\hline
Notation & Definition \\
\hline\hline

\multicolumn{2}{l}{\textsl{Observation}} \\
\hline

\data & Imaging data \\
\datacovmatrix & Data covariance matrix \\

\hline
\multicolumn{2}{l}{\textsl{Model components}} \\
\hline

\model & Image model \\
\convop & Blurring operator \\
\lensingop & Lensing operator \\
\lenspot, \source, \lenslight & Lens potential, source light, lens light \\

\hline
\multicolumn{2}{l}{\textsl{Model parametrization}} \\
\hline

\modelparam & Vector containing all model parameters \\
$\modelparam_{\lenspot}$, $\modelparam_{\source}$, $\modelparam_{\lenslight}$ & Parameters for analytical profiles $\tilde{\lenspot}$, $\tilde{\source}$, $\tilde{\lenslight}$ \\
\lenspotpix & Pixel values of the pixelated potential \\
$\lambda_{\rm\lenspotscalar,bl}$, $\lambda_{\rm\lenspotscalar,bl}$ & Regularization strengths for \lenspotpix \\
$\mat{W}_{\rm\lenspotscalar,bl}$, $\mat{W}_{\rm\lenspotscalar,bl}$ & Regularization weights for \lenspotpix \\

\hline
\multicolumn{2}{l}{\textsl{Optimization \& Inference}} \\
\hline

\loss, \gradloss & Loss function and its gradient \\
\modelparamMAP & MAP solution  \\
$\mat{C}_\modelparam$ & Parameter covariance matrix \\
\hline
\end{tabular}
}
\end{table}

\subsection{Discrete formulation of lensing \label{ssec:strong_lensing}}

From the observation of a strongly lensed source, or data \data, we aim to recover the lens potential \lenspotscalar and the (unlensed) source light \source. Additionally, the lens light \lenslight, which is often partially blended with the lensed features, must also be modeled, either jointly with the lens potential and the source, or carefully subtracted from the data beforehand.

The mapping between an observed angular position \vec{\theta} on the sky and a correpsonding position on the source plane \vec{\beta} before lensing is given by the lens equation
\begin{align}
    \label{eq:lens_eq}
    \vec{\beta} = \vec{\theta} - \vec{\alpha}(\vec{\theta}) \ ,
\end{align}
where the reduced deflection angle $\vec{\alpha}$ is given by the first derivatives of the lens potential
\begin{align}
    \label{eq:defl_field}
    \vec{\alpha}(\vec{\theta}) = \nabla\lenspotscalar(\vec{\theta}) \ .
\end{align}
It is possible to infer the (dimensionless) surface mass density of the lens from the lens potential, called the convergence $\kappa$, which can be computed from the second derivatives of the lens potential, as
\begin{align}
    \label{eq:convergence}
    \kappa(\vec{\theta}) \equiv \frac{\Sigma(\vec{\theta})}{\Sigma_{\rm crit}} = \frac12 \nabla^2 \lenspotscalar(\vec{\theta}) \ ,
\end{align}
where $\Sigma$ is the surface mass density in physical units and $\Sigma_{\rm crit}$ the critical density that depends on the cosmology and the redshifts of the lens and source galaxies. For quoting quantities in physical units (e.g. masses), we assume a flat $\Lambda$CDM model with Hubble constant $\Hc=70$ and matter density at present time $\Omega_{{\rm m},0} = 0.3$.

The lens equation holds for any light ray from the source to the observer, and as such describes the lensing effect in a continuous way. However, the data is pixelated and includes additional nuisance effects due to limiting seeing conditions and instrumental noise. We thus discretize the problem and write \data as
\begin{align}
\label{eq:linear_lens}
    \data = \convop\,\lensingop\,\source + \convop\,\lenslight + \noise \ ,
\end{align}
where \source and \lenslight are vectors holding the true (discretized) surface brightness values of the unlensed source and the lens, respectively. The lensing operator \lensingop encodes a discrete version of Eq.~\ref{eq:lens_eq} that models the lensing effect by mapping source surface brightness values onto the lens plane, based on the lens potential \lenspot (which can also be discretized) through Eq.~\ref{eq:defl_field}. The blurring operator \convop models the seeing conditions by convolving the images of the lens and the lensed source with the point spread function (PSF) of the instrument. Throughout this paper, we assume that the PSF has been modeled beforehand (e.g. from stars in the field) and that it remains constant in Eq.~\ref{eq:linear_lens}. The last term, \noise, represents additive noise, and is usually a combination of instrumental read-out noise (Gaussian noise) and shot noise (Poisson noise that depends on \source and \lenslight). We note that while the model depends linearly on \source and \lenslight, the lensing operator \lensingop depends non-linearly on \lenspot through the lens equation (Eq.~\ref{eq:lens_eq}).

\subsection{Lens modeling and inversion of the lens equation \label{ssec:model_inversion}}

The model defined in Eq.~\ref{eq:linear_lens} cannot easily be inverted to retrieve the lens potential, the unlensed source light, and the lens light. The problem is ill-posed and subject to degeneracies that cannot be mitigated based only on the data: (1) some locations in the source plane do not map to any data pixel (thus are unconstrained), (2) the lens potential is directly constrained only at locations where there are lensed features, (3) the lens light is often blended with these same lensed features. Therefore additional constraints, based on priors on \lenspot, \source and \lenslight, are needed to solve Eq.~\ref{eq:linear_lens} and obtain physically motivated solutions. This is particularly important in situations where the number of model parameters is comparable to the number of pixels in the data, which is a common situation when modeling complex sources or perturbations in the lens potential.

Before specifying the choice of priors, we first simplify the notation by defining the model \model and the corresponding full set of parameters \modelparam that describe \lenspot, \source and \lenslight as
\begin{align}
    \model(\modelparam) \equiv \model(\lenspot, \source, \lenslight) = \convop\,\lensingop\source + \convop\,\lenslight \ .
\end{align}
Inverting Eq.~\ref{eq:linear_lens} comes down to obtaining the set of maximum a posteriori (MAP) parameters \modelparamMAP that maximizes the posterior probability distribution \proba{\modelparam\,|\,\data,\model} of the parameters given the data. From Bayes theorem we have
\begin{align}
\label{eq:posterior_bayes}
    \proba{\modelparam\,|\,\data,\model} = \frac{ \proba{\data\,|\,\modelparam,\model} \proba{\modelparam\,|\,\model} }{ \proba{\data\,|\,\model} } \ ,
\end{align}
where the first term of the numerator is the data likelihood, the second term is the prior, and the denominator is the Bayesian evidence. The evidence does not change \modelparamMAP, but is particularly relevant for comparing different models (through evidence ratios).

In practice, we obtain \modelparamMAP by minimizing a loss function $\loss(\modelparam)$ defined as
\begin{align}
    \label{eq:loss_function}
    \nonumber
    \loss(\modelparam)
    &= - \log\proba{\data\,|\,\modelparam,\model} - \log\proba{\modelparam\,|\,\model} \\
    &\equiv \likelihood{\data\,|\,\modelparam,\model} + \prior{\modelparam} \ ,
\end{align}
where the data-fidelity term \likelihoodsimple is the negative log-likelihood, and the regularization term \priorsimple is the negative log-prior. The MAP solution can then be obtained as
\begin{align}
    \label{eq:map_from_min_loss}
    \modelparamMAP = \argmin_{\modelparam}\ \loss(\modelparam) \ .
\end{align}
The choice of the data-fidelity term is tightly linked to the statistical properties of the data noise \noise, characterized by the covariance matrix of the data \datacovmatrix. The noise is composed of instrumental readout noise and shot noise due to both the flux from the observed target and the sky brightness, which we assume follow Gaussian distributions. The target shot noise is estimated from the ``modeled'' flux \model, as estimating it from the data itself can introduce biases \citep{Horne1986}; therefore we formally have $\datacovmatrix\equiv\datacovmatrix(\modelparam)$ that we implicitly assume throughout the following equations to avoid clutter. Moreover, we assume uncorrelated noise as is usually true for charge-coupled device images, hence $\datacovmatrix$ is a diagonal matrix. A diagonal element of $\datacovmatrix$ for a given data pixel $p$ is given by
\begin{align}
    \label{eq:diag_data_cov}
    \sigma_{\data,p}^2 = \sigma^2_{\rm bkg} + \frac{m_p}{t_{\rm exp}} \ ,
\end{align}
where $\sigma^2_{\rm bkg}$ is the variance of the background noise (readout and sky brightness), $m_p$ is the modeled flux \model at pixel $p$ (in electrons per second) and $t_{\rm exp}$ is the exposure time. The last term of the equation is the Gaussian approximation of the shot noise variance (Poisson noise) due to the lens and source flux.

Under the assumption of Gaussian noise, the data-fidelity term is the $\chi^2$ of the data given the model
\begin{align}
\label{eq:likelihood_chi2}
\nonumber
    \likelihood{\data\,|\,\modelparam,\model} &= \frac12\,\chi^2 \\
    &= \frac12\, \Big[\,\data - \model\left(\modelparam\right)\Big]^\top \datacovmatrix^{-1}\ \Big[\,\data - \model\left(\modelparam\right)\Big] \ .
\end{align}
The priors and corresponding regularization terms depend on the specific choices of parametrization of model components \lenspot, \source and \lenslight. In general, each of these model components can be described by a set of analytical profiles, pixelated profiles, or a combination of both.

\subsection{Analytical and pixelated components}

We model smooth mass and light distributions with a set of analytical profiles. In this case, regularization terms in Eq.~\ref{eq:loss_function} are not explicitly defined, but rather directly encoded in the parametrization of the model \model, which we write as
\begin{align}
    \label{eq:def_analytical_comp}
    \lenspot \equiv \tilde\lenspot(\modelparam_\lenspotscalar), \ \source \equiv \tilde\source(\modelparam_\sourcescalar), \ \lenslight \equiv \tilde\lenslight(\modelparam_\lenslightscalar) \ ,
\end{align}
where $\modelparam_\lenspotscalar$, $\modelparam_\sourcescalar$, and $\modelparam_\lenslightscalar$ are the set of parameters for the analytical profiles $\tilde\lenspot$, $\tilde\source$ and $\tilde\lenslight$ which describe the lens potential, the source and the lens light, respectively.

To describe more complex mass and light distributions that cannot be captured by analytical functions, we rely on pixelated components, where pixel values represent the lens potential, the lens light, or source light (in source plane) at each pixel position. We adopt the following notation for those components
\begin{align}
    \label{eq:def_pixelated_comp}
    \lenspot \equiv \lenspotpix, \ \source \equiv \sourcepix, \ \lenslight \equiv \lenslightpix \ .
\end{align}
Such pixelated components typically imply a much larger number of parameters (i.e., each single pixel value). The model inversion is then highly under-constrained, and the choice of regularization plays a central role in the success of the method in recovering the underlying lens potential and light distributions \citep[see e.g.][]{Vernardos2022}.

\subsection{Multiscale regularization of pixelated components \label{ssec:wavelet_reg}}

We employ a multiscale strategy based on wavelet transforms and sparsity constraints to regularize the pixelated components of the model. While in \citet{Galan2021} we focused on the source model, in this work we consider a pixelated component only in the lens potential. To clarify the relationship between the two works, we first recall the principle of the technique in the context of the source reconstruction, then we apply it to the case of the lens potential.

Our regularization strategy is based on wavelet transforms, which decompose an image into a set of wavelet coefficients organized by spatial scale. Each of these scales is a filtered version of the signal (i.e., same number of pixels) that contains emphasized features at a given spatial scale, similar to a frequency decomposition using the Fourier transform. In \citet{Galan2021}, we use the following regularization term (Eq.~\ref{eq:loss_function}) for the pixelated source component \sourcepix:
\begin{align}
    \label{eq:wavelet_regul_src}
    \priorspec{\source}{\sourcepix} &= i_{\geq0}(\sourcepix) + \lambda_{\sourcescalar}\,\normone{ \weights_\sourcescalar \circ \waveletop^\top\, \sourcepix}  \ .
\end{align}
The first term in the above equation is a positivity constraint\footnote{The indicator function $i_{\geq0}(\,\cdot\,)$ is formally equal to 0 if its argument contains only nonnegative values, and $+\infty$ otherwise.} that enforces pixel values to be nonnegative. The second term combines the $\ell_1$-norm \normone{\,\cdot\,} with the wavelet transform operator $\waveletop^\top$. The effect of the $\ell_1$-norm is to impose a sparsity constraint on wavelet coefficients, which is effectively equivalent to a soft-thresholding\footnote{Soft-thresholding is the proximal operator of the $\ell_1$-norm.} of the coefficients \citep{Starck2015book}. The threshold level depends directly on the hyper-parameter $\lambda_{\sourcescalar}$, which is further adapted to each wavelet scale through the weight matrix $\weights_\sourcescalar$ to efficiently regularize features that span different spatial scales in the source plane (the operation $\circ$ represents the element-wise product). We note that a similar regularization term \priorspec{\lenslight}{\lenslightpix} can also be written for reconstructing the lens light \citep{Joseph2019}.

The success of this regularization strategy thus relies on our ability to correctly estimate the regularization weights $\weights_\sourcescalar$. We follow a data-driven approach that relies on the noise, in the goal to control the statistical significance of the reconstructed source light distribution via $\lambda_{\sourcescalar}$ \citep{Paykari2014}. We achieve this by estimating the standard deviation of the noise in the source plane for each wavelet scale. The details of this procedure are given in \citet{Joseph2019}. The only remaining hyper-parameter is the overall strength of the regularization $\lambda_{\sourcescalar}$, which is a scalar and usually set between 3 and 5 to ensure high enough statistical significance of the reconstructed source light distribution \citep[e.g.][]{Starck2007}.

\begin{figure*}[!tbh]
    \centering
    \includegraphics[width=\linewidth]{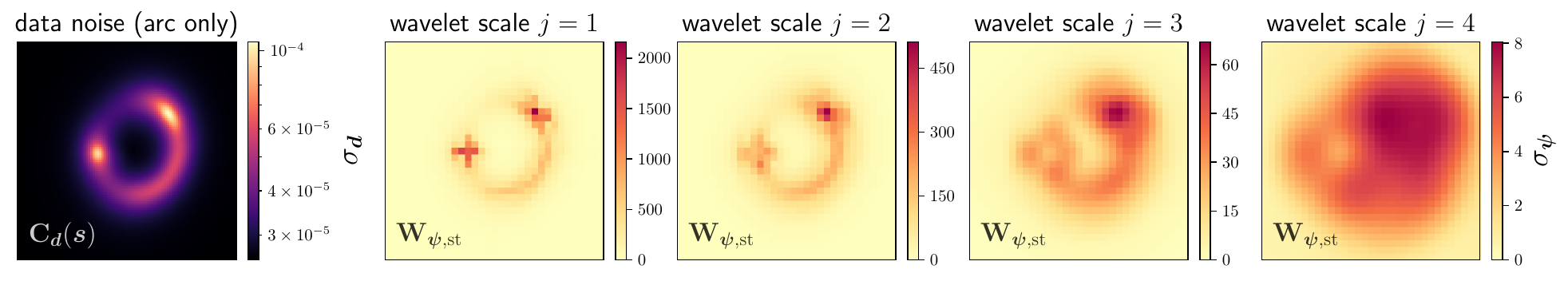}
    \caption{Example maps illustrating the computation of the regularization weights for Eq.~\ref{eq:wavelet_regul_pot}, to properly scale the multiscale regularization of the pixelated lens potential model \lenspotpix. The top-left panel shows the noise standard deviation estimated from a preliminary smooth model of the source, that is the square root of the diagonal elements of \datacovmatrix, without the contribution from the lens light. The remaining panels show the noise standard deviation propagated to each potential pixel for the first four wavelet scales of the starlet transform, that we used as the regularization weights $\weights_{\lenspotscalar,\rm st}$. We see that the standard deviation strongly decreases as a function of the wavelet scale. The resulting maps are similar for the Battle-Lemarié wavelet transform. More details are given Sect.~\ref{ssec:wavelet_reg}.}
    \label{fig:noise_wavelet_pixpot}
\end{figure*}

In this work, we aim to regularize the pixelated potential component \lenspotpix. While in principle, this component can be used to model either the full lens potential or only perturbations to the underlying smooth potential, we focus on the latter case and use \lenspotpix to reconstruct various types of perturbations. The regularization needs to be flexible enough to allow for a variety of different features to be reconstructed, ranging at least from localized subhalos to populations of subhalos and multipolar moments. Therefore, we expect that a multiscale regularization strategy similar to \sourcepix can be applied on \lenspotpix as well. However we do not impose a positivity constraint on the values of \lenspotpix, as negative potential pixels correspond to a local decrease of the potential, relative the smooth potential component.

The full regularization term for \lenspotpix is
\begin{align}
    \label{eq:wavelet_regul_pot}
    \nonumber
    \priorspec{\lenspot}{\lenspotpix} &= \lambda_{\lenspotscalar,\rm st}\,\normone{\weights_{\lenspotscalar,\rm st} \circ \starletop^\top\, \lenspotpix} \\
    &+ \lambda_{\lenspotscalar,\rm bl}\,\normone{\weights_{\lenspotscalar,\rm bl} \circ \battlelemarieop^\top\, \lenspotpix}  \ ,
\end{align}
where we use two different wavelet transforms: the starlet transform ($\starletop^\top$) and the Battle-Lemarié wavelet transform of order 3 ($\battlelemarieop^\top$). As in Eq.~\ref{eq:wavelet_regul_src}, the scalars $\lambda_{\lenspotscalar,\rm st}$ and $\lambda_{\lenspotscalar,\rm bl}$ are the regularization strengths, and the elements of the matrices $\weights_{\lenspotscalar,\rm st}$ and $\weights_{\lenspotscalar,\rm bl}$ are the regularization weights. The starlet transform is the transform used for modeling the source light distribution in \citet{Galan2021}, and is well-suited for the reconstruction of multiple locally isotropic features at different spatial scales. The Battle-Lemarié transform is introduced to reduce the appearance of spurious isolated potential pixels, inspired by a similar use for mass-mapping studies from weak lensing observations \citep{Lanusse2016}.

Similarly to the regularization of the source, we aim to find the regularization weights $\weights_{\lenspotscalar,\rm st}$ and $\weights_{\lenspotscalar,\rm bl}$ based on the noise in the data. The situation is however more complicated than for the pixelated source reconstruction discussed in Sect.~\ref{ssec:wavelet_reg}, because the relation between the lens potential and the lensed source light distribution is highly nonlinear due to the lens equation. Nevertheless, in the limit of small perturbations $\delta\lenspot$ to the smooth potential \lenspot, it is possible to linearize the lens equation and relate changes in the lens potential to changes in the lensed light distribution of the source. We use this first-order treatment for estimating the covariance---more specifically the standard deviation---in the lens potential, and scaling the regularization strengths accordingly.

A first-order Taylor expansion of the lens equation around $\vec\beta$ leads to the following equation \citep{Blandford2001,Koopmans2005}
\begin{align}
\label{eq:deta_d_delta_psi_func}
    \delta\data(\vec\theta) \approx -\nabla_\beta\, \source(\vec\beta)\cdot\nabla_\theta\,\delta\lenspot(\vec\theta) \ ,
\end{align}
where residuals $\delta\data \equiv \data - \tilde{\model}(\lenspot, \source)$ are based on a preliminary model $\tilde{\model}$ that does not include any perturbations $\delta\lenspot$ in the lens potential. Gradients are computed with respect to the coordinates indicated by the subscripts $\beta$ and $\theta$ (defined in Eq.~\ref{eq:lens_eq}). The above equation provides a linear relation that connects individual pixels in potential space to individual pixels in data space, which we can use to propagate noise levels for tuning the regularization strengths. We give all the remaining details of the computation in Appendix~\ref{app:sec:regularization_weights}. The resulting weight matrices $\weights_{\lenspotscalar,\rm st}$ and $\weights_{\lenspotscalar,\rm bl}$ contain the standard deviation of the noise for each wavelet scale in the lens potential. We show an example of weights with respect to the starlet transform (i.e., $\weights_{\lenspotscalar,\rm st}$) in Fig.~\ref{fig:noise_wavelet_pixpot}.

As for the source, the remaining hyper-parameters are the two regularization strengths $\lambda_{\lenspotscalar,\rm st}$ and $\lambda_{\lenspotscalar,\rm bl}$, which directly control the statistical significance of the starlet and Battle-Lemarié wavelet coefficients, respectively. These scalars are set in practice between 3 and 5, depending on how strongly certain features need to be regularized.

\subsection{Optimization with differentiable programming \label{ssec:autodiff}}

So far we have only defined how the different model components (\lenspot, \source, \lenslight) can be parametrized and cast into an optimization problem. However, combining components which are described either with analytical profiles or on pixelated grids is a challenging task, as the standard methods to minimize the loss function can be fundamentally different in each case. With analytical model components, the MAP solution \modelparamMAP can be approached via stochastic algorithms such as particle swarm optimization \citep[PSO,][]{Kennedy2001_PSO}. With wavelet-regularized pixelated components (Eqs.~\ref{eq:wavelet_regul_src} and \ref{eq:wavelet_regul_pot}), convergence to \modelparamMAP usually requires the use of carefully chosen iterative algorithms relying on the formalism of proximal operators to apply constraints such as sparsity and positivity to the solution \citep[so-called proximal splitting algorithms, see e.g.][]{Starck2015book}.

In \citet{Galan2021}, we used a hybrid scheme that first optimizes the smooth lens potential described analytically using a PSO, followed by a source reconstruction step (at fixed lens potential parameters) using an iterative proximal algorithm. However, this strategy does not scale well with model complexity. Each additional pixelated model component would require similar hybrid schemes, which rapidly become inefficient at converging to the MAP solution. In addition, estimating the joint posterior distribution of model parameters using traditional sampling techniques (e.g. Markov chain Monte Carlo, MCMC) becomes computationally too expensive.
In this work, we overcome this issue by implementing a fully differentiable loss function, meaning that we can obtain its full gradient and higher order derivatives analytically. This allows us to simultaneously optimize all parameters, both analytical parameters and individual pixel values, using robust gradient descent algorithms which remain efficient even in a large parameter space. This naturally replaces proximal algorithms that are usually necessary to solve for pixelated components, and leads to a self-consistent combination of both analytical and pixelated model components.

Gradient descent optimization guarantees convergence to a minimum of the loss function, which is typically not the case for stochastic optimization algorithms that do not use the gradient (or higher order derivatives) of the loss function. However there is still no definitive guarantee to converge to the global minimum (i.e., \modelparamMAP), which can depend on the parameter initial values. It is possible to address this limitation by using a multistart gradient descent optimization, which runs the same minimization multiple times for different parameter initializations \citep[e.g.][]{Gu2022}. We leave this optimization improvement to future work.

We use the automatic differentiation Python library \jax \citep{jax2018github} to construct a fully differentiable model and loss function. Wavelet transforms are implemented using convolutions, making them straightforwardly differentiable (as in convolutional neural networks). We also take advantage of efficient compilation features of \jax to speed up computations during evaluation of the loss function and its derivatives.
All of our modeling methods and algorithms are implemented in the Python software package \herculens, which we make publicly available\footnote{\url{https://github.com/austinpeel/herculens}}. The code structure and part of the modeling routines of \herculens are based on the open-source modeling software package \lenstro \citep{Birrer2018lenstro,Birrer2021lenstro}. All modeling and analysis scripts are also publicly available\footnote{\texttt{\url{https://github.com/aymgal/wavelet-lensing-papers}}}.

\subsection{Estimating the parameter covariance matrix \label{ssec:fisher_uncertainties}}

Estimating parameter uncertainties and covariances is crucial to reliably interpret the model that corresponds to the MAP solution. Sampling techniques such as MCMC, which draw samples from the full posterior distribution function (Eq.~\ref{eq:posterior_bayes}), are often used for this purpose. However, for large parameter spaces, such stochastic techniques become inefficient, especially when parameters depend non-linearly on each other, as is the case in lensing.

In this regime, Hamiltonian Monte Carlo sampling \citep[HMC, introduced by][for a review]{Duane1987,Neal2011} is particularly efficient as each new sample is drawn based on the gradient of the loss function, resulting in high acceptance rates. While individual HMC steps might be more expensive to compute, the number of samples required for a reliable estimation of the parameters' posterior distributions is largely reduced compared to stochastic techniques such as MCMC, which tend to deliver noisier distributions. Moreover, we note that gradient-informed nested sampling for calculating the Bayesian evidence also exists \citep[see e.g.][]{Albert2020}, which is an important tool for model comparison (although not explored in this work).

In addition to HMC sampling, we also explore the possibility of using the Fisher information matrix (FIM) and its relation to the second-order derivatives of the loss function to estimate the parameters' covariance matrix. This is usually referred to as a Fisher analysis, which has been used in various studies including the forecast of constraints on cosmological parameters \citep[e.g.][]{Philcox2021}, the mapping of large-scale structures \citep[e.g.][]{Abramo2012}, the analysis of gravitational waves \citep[e.g.][]{Belgacem2019}, and the substructure power-spectrum from lensing \citep{CyrRacine2019}.

Second-order partial derivatives of the loss function \loss define the Hessian matrix, which reflects the local shape of the loss function at any point of the parameter space. Each entry of the Hessian matrix \hessianloss is defined as
\begin{align}
    \label{eq:def_hessian}
    \mat{H}_\loss(\modelparam) = \left\{ \frac{\partial^2\loss(\modelparam)}{\partial\modelparamsgl_i\,\partial \modelparamsgl_j} \right\} \ ,
\end{align}
where the indices $i$ and $j$ indicate different model parameters from the entire parameter set \modelparam. The ``expected'' Fisher information matrix, denoted as \expfim, is defined as
\begin{align}
    \label{eq:def_exp_fisher}
    \expfim(\modelparam) = \mathbb{E}\Big[\hessianloss(\modelparam)\Big] \ ,
\end{align}
where $\mathbb{E}$ is the expectation operator over the distribution of all the realizations of the data for a fixed set of parameters $\modelparam$ (i.e., $\proba{\data\,|\,\modelparam}$).
In practice, however, we cannot compute the expected value $\expfim(\modelparam)$, because we only have access to a single observation \data, with a specific realization of the noise. Instead, we can compute the ``observed'' FIM, $\obsfim(\modelparam)$, evaluated at the MAP solution
\begin{align}
    \label{eq:def_obs_fisher}
    \obsfim(\modelparamMAP) = \hessianloss\left(\modelparamMAP\right) \ .
\end{align}
The FIM can then be used to approximate the covariance matrix of the MAP solution through matrix inversion:
\begin{align}
    \label{eq:cov_from_fisher}
    \covmatrix_\modelparam \approx \obsfim(\modelparamMAP)^{-1} \ .
\end{align}
This covariance directly gives a lower bound on the uncertainty for each parameter\footnote{This is more formally called the Cramér-Rao bound, which states that the variance of an unbiased parameter estimator is at least as high as the inverse of the Fisher information \citep[e.g.][]{Cramer1999}.}. We note that this (first-order) approximation of $\covmatrix_\modelparam$ becomes exact if the loss function locally behaves quadratically or, equivalently, follows a Gaussian distribution. We expect this to be the case for simple, smooth models described with analytical profiles, as shown in \citet{Vernardos2022}. We find that while for fully analytical models we can rely on the above first-order approximation of parameter covariance matrix, a sampling-based exploration of the highly nonlinear parameter space using HMC is warranted.

\section{Experimental setup \label{sec:experimental_setup}}

\begin{figure*}[!tbh]
    \centering
    \includegraphics[width=\linewidth]{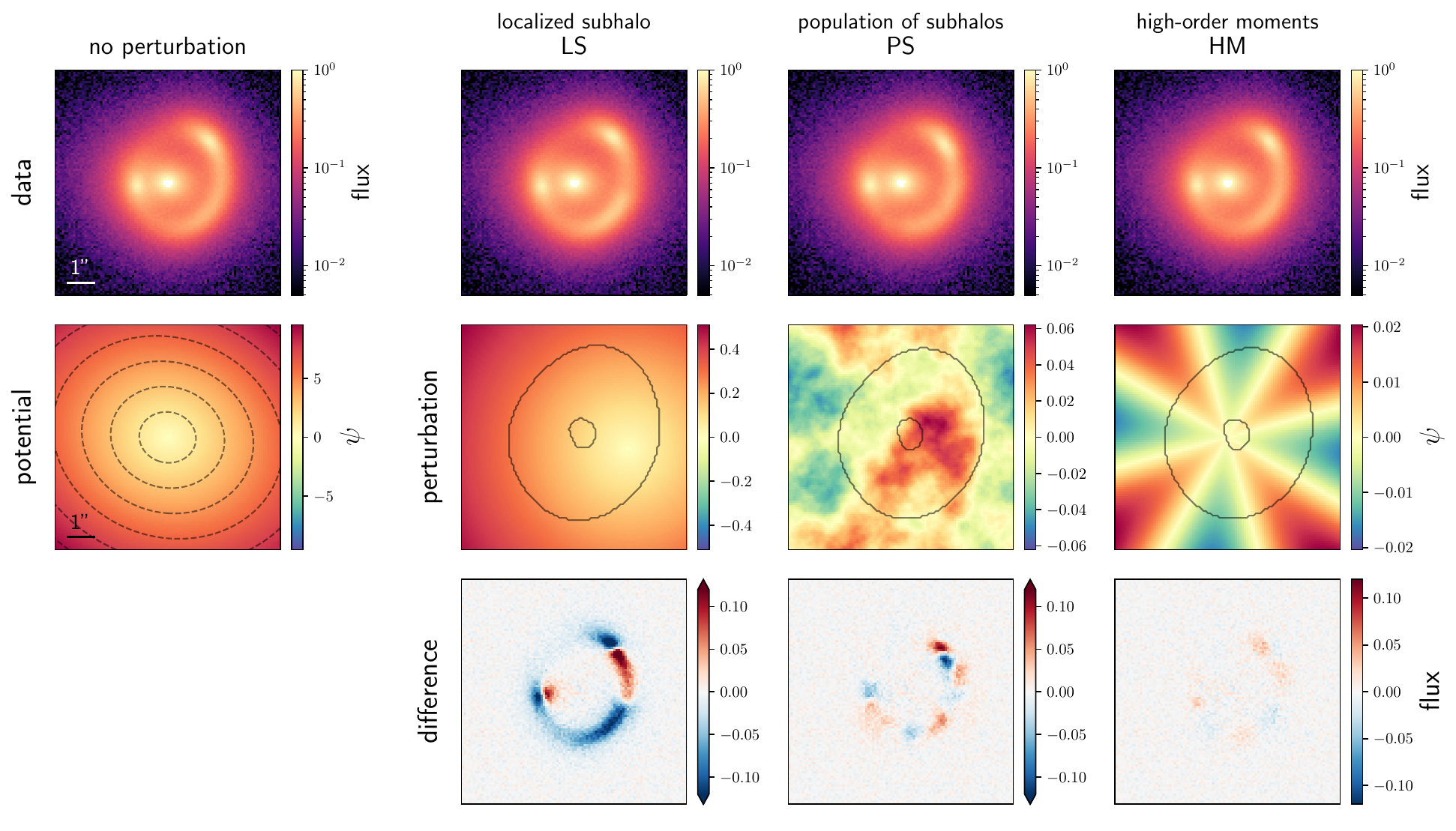}
    \caption{Overview of the simulated HST observations. The left-most column shows the simulated data without perturbations to the smooth lens potential. The dashed lines in the bottom-left panel are isopotential contours. The remaining columns show the mock data used in this work (top row), the three different input potential perturbations (middle row), and the difference between the unperturbed and perturbed lenses (bottom row). The range (min and max) of the perturbations varies according to the different nature of the perturbers. The solid lines enclose the region where the S/N of the lensed source is higher than 5. The isopotential contours are almost indistinguishable between the unperturbed and perturbed cases, therefore we omit them.}
    \label{fig:data_summary}
\end{figure*}

We test our method on mock Hubble Space Telescope (HST) observations that include realizations of the three perturbation categories introduced in Sect.~\ref{sec:intro}: a localized DM subhalo, a population of DM subhalos, and higher-order multipoles. We simulate these perturbing fields based on physical considerations and results of previous works (detailed in the sections below), which result in each case in very different levels of perturbations when compared to the underlying smooth lens potential.

We focus on the strong lensing of a source with smooth surface brightness, resembling an early-type galaxy lensed by a foreground early-type galaxy \citep[so-called EEL,][]{Oldham2017}. We simulate both the source and the lens light as single elliptical Sérsic profiles, as early-type galaxies are often well-fitted by these profiles \citep[e.g.][]{Shu2017}. The smooth lens potential is described by a singular isothermal ellipsoid (SIE) for the main deflector, embedded in an external shear to simulate the net influence of neighboring galaxies. All analytical profiles used in this work are defined in Appendix~\ref{app:sec:analytical_models}.

To create the mock data, we use an algorithm that is entirely independent of our modeling code, in order to prevent the occurrence of artificially advantageous minima during optimization. This also has the advantage to closely mimic a real-world situation, as the data never exactly corresponds to any model generated by the modeling code itself. We use the software package \molet \citep{Vernardos2021} to simulate typical observations of strongly lensed galaxies as observed with HST and the Wide Field Camera 3 (WFC3) instrument, in the near infrared (F160W filter). The pixel size is $0\farcs08$, and the field of view is $8''\times8''$. In this work we do not explore effects due to incorrect PSF modeling, hence for simplicity we use a gaussian PSF with $0\farcs3$ FWHM. This results in a simulated data set that is sufficiently realistic to evaluate our method. The first column of Fig.~\ref{fig:data_summary} shows the simulated image without perturbations to the smooth potential. The remaining three columns show the different perturbation cases to which we apply our method. Full details of all instrumental settings and input model parameters are listed in Appendix~\ref{app:sec:sim_settings}.

\subsection{Localized subhalo (LS) \label{ssec:lens_LS}}

We simulate a single localized DM subhalo, with a spherical isothermal profile (SIS) and Einstein radius $\theta_{\rm E,halo}=0\farcs07$ as input perturbing potential (for reference, the Einstein radius of the main deflector is $1\farcs6$). For simplicity, we assume that the subhalo is in the same redshift plane as the main lens, although it can be located at any redshift (through a simple scaling of its mass). We assume typical lens and source redshifts of EEL lenses $\zd = 0.3$ and $\zs = 0.7$ respectively \citep{OldhamAuger2018}, from which we get a mass of $10^9\ {\rm M}_{\odot}$ within the SIS Einstein radius, which is comparable to previous dark halo detections based on HST observations \citep{Vegetti2010}. The resulting simulated observation is shown in the second column of Fig.~\ref{fig:data_summary}. The mean perturbation level, computed within the region containing the lensed arcs, is $6.4\%$ relative to the smooth potential.

\subsection{Population of subhalos (PS)}

We follow recent work and simulate the net effect of a population of DM subhalos along the line-of-sight with a gaussian random field (GRF) \citep{ChatterjeeKoopmans2018,Bayer2018,Vernardos2020}. GRF perturbations are random fluctuations that have a Fourier power-spectrum following a power-law. We parametrize this power-law relation as
\begin{align}
    \label{eq:def_grf}
    {\rm PS}\big(\psi_{\rm GRF}\big)(k) = C_\beta \, \sigma^2_{\rm GRF}\,k^{-\beta_{\rm GRF}} \ ,
\end{align}
where $k$ is the wavenumber, $\beta_{\rm GRF}$ is the power-law slope, and $C_\beta$ is a normalizing factor that depends on $\beta_{\rm GRF}$ and the size of the field of view, which ensures that $\sigma^2_{\rm GRF}$ is the variance of the GRF \citep[for the exact formula, see][]{Chatterjee2019_thesis}. The power-spectrum is then converted to a specific random realization of direct-space perturbations using the inverse Fourier transform. The value of $\beta_{\rm GRF}$ determines  the distribution of power at each length scale: a large value leads to extended and smooth variations, whereas a small value creates a large number of localized and grainy structures.

Typical ranges for $\sigma^2_{\rm GRF}$ and $\beta_{\rm GRF}$ have been explored in the literature and are justified in \citet{Chatterjee2019_thesis} and \citet{Vernardos2020}. In these works, the authors explored ranges $\sigma^2_{\rm GRF} \in [10^{-5},10^{-2}]$ and $\beta_{\rm GRF} \in [3,8]$. Additionally, \citet{Bayer2018} excluded GRF variance larger than $\sigma^2_{\rm GRF}=10^{-2.5}$, based on HST observations of the strong lens system SDSS\,J0252$+$0039. Hence, we set $\sigma^2_{\rm GRF}=10^{-3}$ and $\beta_{\rm GRF}=4$, such that it leads to a GRF that is not unphysically large, and contains both small and large-scale features. The specific GRF realization and the corresponding simulated observation are shown in the third column of Fig.~\ref{fig:data_summary}. The corresponding relative mean perturbation level is $0.83\%$.

\subsection{Higher-order multipoles (HM)}

We introduce higher-order deviations to the smooth potential as a multipole of order 4 (octupole). We use the same definition of multipole as in \citet{VandeVyvere2022_TDC7}
\begin{align}
    \label{eq:def_multipole}
    \psi_{\rm multipole}(\vec{\theta}) = \frac{r}{1-m^2}\,a_m\cos\!\left(m\phi - m\phi_m\right) \ ,
\end{align}
where $r$ and $\phi$ are polar coordinates transformed from \vec{\theta}. In our case, we fix the multipole order to $m=4$ (octupole), and the orientation $\phi_4$ to be aligned with the main axis of the SIE component of the smooth potential (in this case the full potential becomes more disky). The octupole strength is set to $a_4=0.06$, which corresponds to the high end of the distribution found by \citet{Hao2006}, from isophote measurements over a large sample of elliptical and lenticular galaxies in SDSS data. The resulting simulated observation is shown in the last column of Fig.~\ref{fig:data_summary}. The corresponding relative mean perturbation level is $0.16\%$.

\section{Modeling the full lens potential \label{sec:modeling_results}}

\subsection{Baseline model, parameter optimization and sampling}

We model the simulated data set presented above, with the goal of retrieving the full lens potential, including the perturbations. The lens potential is modeled as $\lenspot = \tilde\lenspot(\modelparam_\lenspotscalar) + \lenspotpix$. The smooth component $\tilde\lenspot$ is parameterized as a SIE and external shear, and the a priori unknown perturbations are captured in the pixelated component \lenspotpix, regularized with sparsity and wavelets as discussed in Sect.~\ref{ssec:wavelet_reg}. The deflection angles at each position in the image plane are computed based on bicubic interpolation of \lenspotpix. We use bicubic instead of bilinear interpolation in order to compute the surface mass density (Eq.~\ref{eq:convergence}) corresponding to the pixelated model, as it requires second-order spatial derivatives of the potential (these derivatives are always zero with bilinear interpolation). We model the surface brightness of the source using a Sérsic profile, that is $\source = \tilde\source(\modelparam_\sourcescalar)$. This modeling choice means that we assume an accurate knowledge of the underlying shape of the source galaxy.

The lens light profile is modeled only once with a Sérsic profile for one of the system, assuming the other model components are known, then it is then fixed during the rest of the modeling. Fixing the lens light is not identical to subtracting it from the observation, as it still contributes to the noise model and reduces the contrast of the lensed features. We assume the lens light centroid is a good tracer of the mass centroid, which is a realistic scenario for fairly isolated lens galaxies \citep[see e.g.][]{Shajib2019}. Therefore, we join the center of the SIE profile to that of the lens light profile. We note that it is however outside the scope of this work to assess the impact of inaccurate lens light modeling.

We model instrumental and seeing effects by assuming perfect knowledge of the PSF, background noise level (read-out noise and shot noise from sky brightness), however we estimate the shot noise from the modeled lens and source light distributions to estimate the diagonal of the data covariance matrix following Eq.~\ref{eq:diag_data_cov}.

When optimizing only analytical profile parameters ($\modelparam_\lenspotscalar$, $\modelparam_\sourcescalar$)---which we do before including pixelated components in the model---, we find that the quasi-Newton optimization method BFGS\footnote{\url{https://docs.scipy.org/doc/scipy/tutorial/optimize.html}} \citep{Nocedal2006} is sufficient to reach convergence to the MAP solution. However, the optimization of both analytical and pixelated model components is more challenging, as the dynamic range of analytical and pixel parameters can vary significantly during optimization according to their impact on the loss function and the different regularization terms. Therefore, in this case, parameter updates are performed using the adaptive gradient descent algorithm \textsc{AdaBelief} \citep{Zhuang2020_AdaBelief}, which is extremely efficient for optimizing a large number parameters (typically as large as for convolutional neural networks). The initial learning rate is set in order to obtain a smooth decrease of the loss function until convergence, coupled with an exponential decay of the learning rate. We use the optimization library \textsc{Optax} \citep{optax2020github} that implements the \textsc{AdaBelief} algorithm.

For fully analytical models, we find that using the fast estimation from the FIM leads to a parameter covariance matrix almost indistinguishable from the one obtained via sampling methods. Therefore we only rely on the FIM for these simple models. However, for the more complex models that include a pixelated component, we find that HMC sampling of the parameter space is warranted to obtain reliable estimates of the posterior distributions. We use the python package \textsc{BlackJAX}\footnote{\url{https://github.com/blackjax-devs/blackjax}} that provides an implementation of HMC well-integrated with \jax, that we run using the ``No U-Turn Sampler'' algorithm \citep[NUTS,][]{Hoffman2011} to dynamically adapt the step size, and their ``Stan's adaptation window'' feature to improve sampling efficiency.

Our fiducial model is defined with a \lenspotpix pixel size set to three times the data pixel size, leading to a resolution of $0\farcs24$. The choice of pixel size is based on preliminary models comparing the best-fit reconstructions and model residuals for different \lenspotpix resolutions. While the residuals do not vary significantly for a pixel scale between 4 and 1.5, best-fit reconstructions obtained with pixel scales larger than 3 display artifacts at the scale of individual pixels. Although most of these artifacts can be efficiently reduced using our multiscale regularization strategy by increasing further the regularization strength for small wavelet scales, we find that using a larger number of model parameters for only marginal improvements in terms of residuals is not necessary for this work. We refer to Appendix~\ref{app:sec:choice_pixel_scale} for further discussion on the choice of pixel scale for \lenspotpix. The total number of parameters for our fiducial model is thus $1101$ ($1089$ for \lenspotpix, $5$ for $\modelparam_\lenspotscalar$ and $7$ for $\modelparam_\sourcescalar$), jointly optimized and constrained by $10^4$ data pixels.

The regularization strengths for \lenspotpix are set to $\lambda_{\lenspotscalar,\rm st}=3$ and $\lambda_{\lenspotscalar,\rm bl}=4$, that is in the range of values discussed in Sect.~\ref{ssec:wavelet_reg}. We use a $1\sigma$ higher strength for the Battle-Lemarié regularization in order to penalize more the appearance of spurious pixels in the solution.

\subsection{Modeling the perturbations only}

\begin{figure*}[!tbh]
    \centering
    \includegraphics[width=0.8\linewidth]{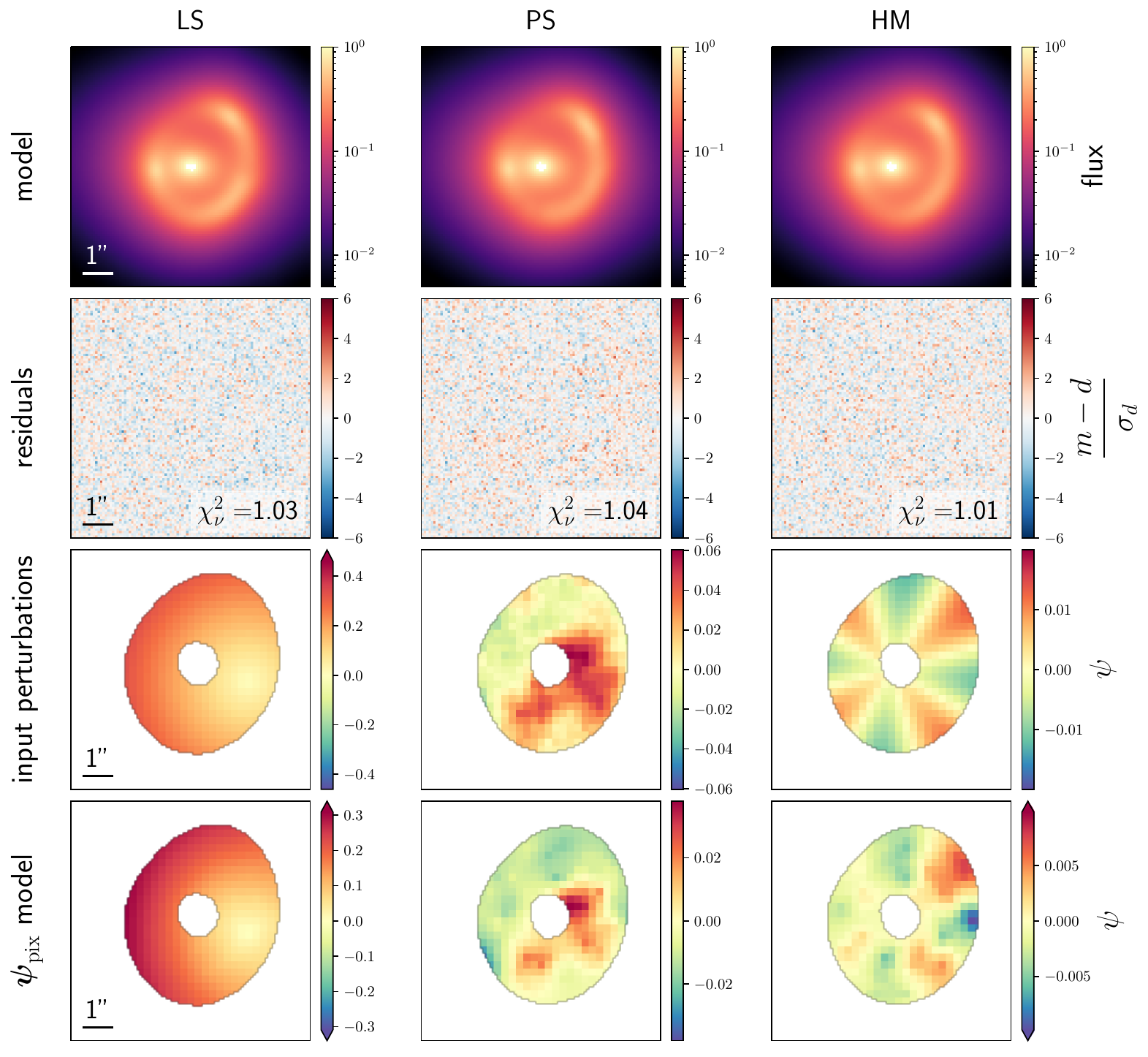}
    \caption{Best-fit pixelated potential reconstruction of our simulated data set, assuming all other model components are perfectly known. Each column represents the different types of perturbations considered in this work. \textit{From top to bottom}: simulated data, image model, normalized residuals, input perturbations, \lenspotpix model. The outlined annular region on the potential panels corresponds to the solid lines in the panels of Fig.~\ref{fig:data_summary}, and is where the S/N of the lensed source is higher than 5. Perturbations outside this area are cropped only to ease the visual comparison with the input perturbations (there is no cropping or masking during modeling). Additionally, for lens LS only, the minimum value of the perturbations is subtracted from the model, again to ease the comparison with the input SIS profile (this does not affect the lensing observables). An animated visualization of the gradient descent optimization is available \href{https://obswww.unige.ch/~galanay/animations/}{on this link}.}
    \label{fig:model_summary_pix_pot_ideal}
\end{figure*}

We first reconstruct the perturbations in the idealized case where all smooth components are perfectly known and fixed. This unrealistic scenario allows us to assess the best level of perturbations that can be recovered, given the quality of the data set. As the source and the smooth potential are fixed, the regularization weights $\mat{W}_{\lenspotscalar, \rm st}$ and $\mat{W}_{\lenspotscalar, \rm bl}$ can be precomputed and kept constant throughout the gradient descent optimization.

The resulting models, corresponding to the MAP parameters \modelparamMAP, are shown in Fig.~\ref{fig:model_summary_pix_pot_ideal}. We also note that the pixelated model \lenspotpix is expected to differ from the input by a uniform offset, as a uniform value in the potential corresponds to zero deflection of light rays (Eq.~\ref{eq:defl_field}). This constant offset is thus never constrained by the data alone (and depends on the initialization). Hence, for better visualizing the reconstruction for lens LS (for which we know that the input potential is positive), we shift the \lenspotpix model such that the minimum pixel value displayed on the figure is zero. We do not add such an offset to our reconstructions for lenses PS and HM shown in Fig.~\ref{fig:model_summary_pix_pot_ideal}, as the underlying perturbations have roughly zero-mean.

Overall the characteristic features of each type of perturbations are well recovered. For lens LS, both the subhalo position and the shape of the underlying SIS profile are captured by the model. For lens PS, the clear over-density region on the bottom right part of the arc is recovered, although the correspondence with the input perturbations is less than for lens LS. For lens HM, the reconstruction displays imprints of azimuthal periodicity between over- and under-density regions, despite the low level (0.2\%) of perturbation. Lastly, we notice that the amplitude of the modeled perturbations is systematically lower, by a factor of $\sim$2, than the input perturbations. We note that a similar result can be seen in some of the pixelated reconstructions of \citet[][see their fig.~8]{Vernardos2022}.

\subsection{Modeling the full potential and the source \label{ssec:full_potential_modeling}}

\begin{figure*}[!tbh]
    \centering
    \includegraphics[width=0.8\linewidth]{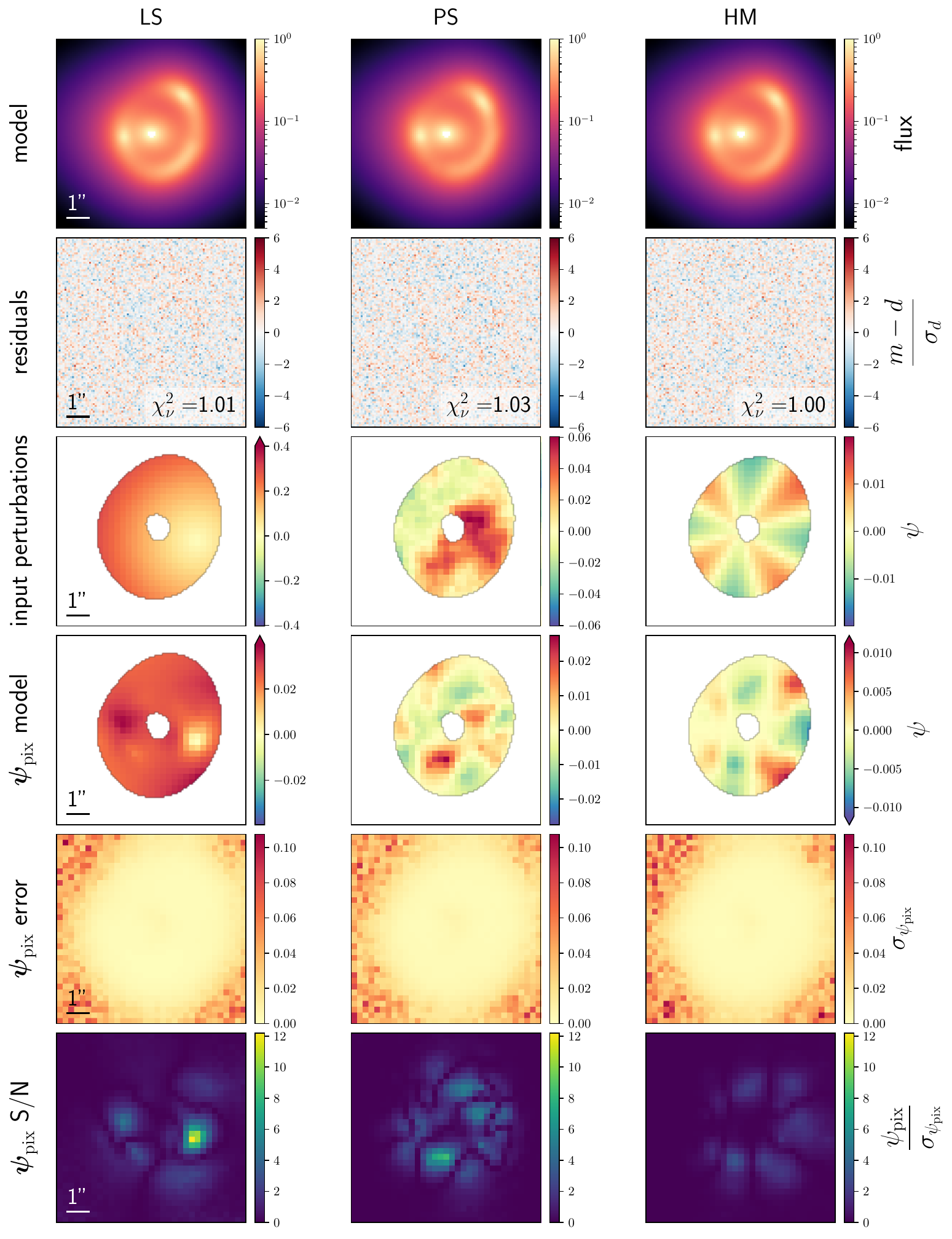}
    \caption{Best-fit models on the simulated data set, after full modeling of the lens potential and the source. The first four rows are as in Fig.~\ref{fig:model_summary_pix_pot_ideal}. The last two rows show, in order: the standard deviation for each \lenspotpix pixel obtained from HMC sampling of the posterior, and the S/N maps that correspond to the absolute value of the modeled potential divided by the standard deviation.}
    \label{fig:model_summary_pix_pot_real}
\end{figure*}

Our method is then applied to the more realistic situation in which the full lens potential and the source light are unknown as well. We do so by uniformly modeling the simulated data set using a series of steps involving gradient descent optimization. After each step, the model degrees of freedom are gradually increased, in order to prevent the optimizer to be trapped in local minima. These optimization steps are as follows.

Firstly, the pixelated component \lenspotpix of the lens potential is initialized to zero (in practice to $10^{-8}$, to prevent gradients to be evaluated at zero, see also Appendix~\ref{app:sec:l1_diff}) and kept fixed. The initial model is thus fully smooth, with corresponding parameters $\modelparam_\lenspot$ and $\modelparam_\source$. We use a multistart gradient descent \citep[as advocated by][]{Gu2022} with 30 runs, for which we verified that it leads to convergence to the global minimum.

Secondly, the pixelated potential component is released and regularization strengths are deliberately set to large values, such that only the most significant potential pixels enter the solution. The regularization weights are computed based on the smooth model from the previous steps, and fixed throughout the gradient descent. This prevents the pixelated component from fitting all model residuals from the previous step, which can strongly bias the recovered perturbations. We find that setting $\lambda_{\lenspot,\rm st} = 10$ and $\lambda_{\lenspot,\rm bl} = 20$ leads to an intermediate solution that contains the main features of the perturbations, reducing slightly model residuals but preventing the model from getting trapped in a spurious minimum. All model parameters ($\modelparam_\lenspot$, $\modelparam_\source$ and \lenspotpix) are simultaneously optimized.

Thirdly, the regularization strengths are set to their fiducial values ($\lambda_{\lenspot,\rm st} = 3$, $\lambda_{\lenspot,\rm bl} = 4$), and regularization weights are recomputed based on the previous models. All model parameters are simultaneously further optimized.

Lastly, we perform HMC sampling to estimate the error and posterior distributions for all parameters. We find that, thanks to the gradient information, only 400 samples (after a warm-up phase of 100 samples) are necessary to obtain well-sampled posteriors. For posterior distributions that are have multiple modes, which is not the case in our models, the number samples should be increased.

In Fig.~\ref{fig:model_summary_pix_pot_real} we summarize, for each lens, the models corresponding to the best-fit parameters \modelparamMAP. The panels are identical to those of the ideal case (Fig.~\ref{fig:model_summary_pix_pot_ideal}), although we added two additional panels to help the interpretation of the results. The first of these panels is the standard deviation for each \lenspotpix pixel, based on the HMC samples, which can be interpreted as an error map for the best-fit model. These error maps are almost identical for each modeled system, but are necessary to compute the second additional panel, which shows a measure of the signal-to-noise (S/N) of the reconstruction. We define the S/N for \lenspotpix as the absolute value of the best-fit model divided pixel-wise by the standard deviation. These maps reveal the regions where the reconstruction is the most statistically significant.

For lens LS, the position of the subhalo is clearly and accurately recovered. However, the overall shape of the underlying smooth subhalo profile is not recovered equally well as in the ideal case. We also observe some features on the other side of the Einstein ring, although those are less significant than at the subhalo position. The S/N map clearly shows that the feature at the position of the subhalo has a higher likelihood compared to other features present in the solution.

For lens PS, we see a relatively good agreement with the input field, where both over- and under-dense regions are recovered. However, again, the amplitude of the perturbation is well below the input value, by a factor between 1.5 and 3. Contrary to the reconstruction of the ideal case, the over-dense region on the lower part of the ring is now the most prominent. This feature is well aligned with a lower-intensity region of the arc, which can partly explain why the model favors a correction to the smooth potential at this location. The S/N map confirms the significance of this feature. All the reconstructed features have a S/N of approximately 3 and above. This system is arguably the most complex to model, as the perturbations affect the lensed features at many different locations with different strengths and orientations, which is likely to translate to a larger set of possible solutions.

For lens HM, the reconstruction reveals a clear azimuthal periodicity, with no regions significantly different from the input perturbations. Interestingly, the reconstruction is very similar to the ideal case, despite the input level of perturbations (relative to the smooth potential) being lower than in the other systems. Similar to lens PS, we notice a region where the model over-predicts a correction to the smooth potential. This region is better revealed in the S/N map and corresponds to the same low-flux region of the arc. This is not surprising as lower S/N in the data leads to looser constraints on the model parameters. For this lens we also notice that the model does not predict any perturbation at the position of left-most image of the source. At this location, which is closer to the lens center compared to other parts of the arc, the input perturbation level is lower while the data noise is higher (shot noise), which leads to almost no detection of perturbations.

\subsection{Effect of the regularization strength \label{ssec:effect_lambda}}

\begin{figure*}[!tbh]
    \centering
    \includegraphics[width=0.95\linewidth]{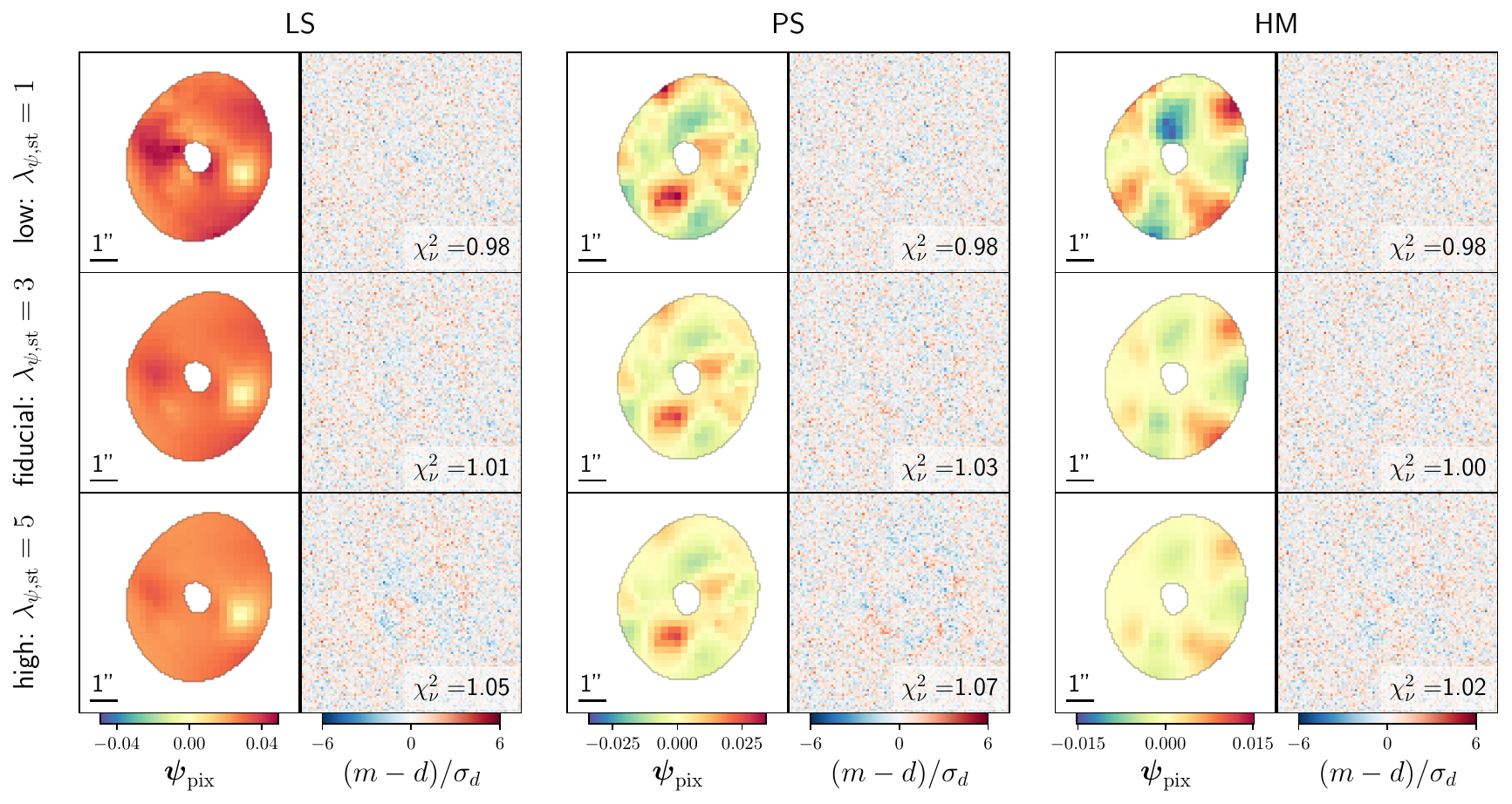}
    \caption{Reconstructed potential perturbations for different values of the regularization strength. The first row corresponds to a low regularization strength with $\lambda_{\lenspotscalar,\rm st}=1$. The second row shows our fiducial models of Fig.~\ref{fig:model_summary_pix_pot_real}, with $\lambda_{\lenspotscalar,\rm st}=3$. The bottom row corresponds to a high strength with $\lambda_{\lenspotscalar,\rm st}=5$. In all cases, the strength for the Battle-Lemarié wavelet ($\lambda_{\lenspotscalar,\rm bl}$) is set $1\sigma$ higher than for the starlet ($\lambda_{\lenspotscalar,\rm st}$) to penalize isolated spurious features in the reconstructed. The main features of the perturbations are recovered in all panels, so the model is robust against reasonable changes in regularization strengths. A too low regularization strength leads to over-fitting ($\chi^2_\nu<1$) whereas a higher regularization strength leads to a smoother model, as expected.}
    \label{fig:model_different_lambdas}
\end{figure*}

The models presented in the previous section correspond to well-motivated but fixed strengths for the regularization of the pixelated potential component. We investigate the effect of the regularization strength on the reconstructed perturbations by running the same modeling procedure with different values of $\lambda_{\lenspotscalar,\rm st}$ and $\lambda_{\lenspotscalar,\rm bl}$. Compared to our fiducial models with $\{\lambda_{\lenspotscalar,\rm st}=3, \lambda_{\lenspotscalar,\rm bl}=4\}$, we define a low regularization case $\{\lambda_{\lenspotscalar,\rm st}=1, \lambda_{\lenspotscalar,\rm bl}=2\}$ and a high regularization case $\{\lambda_{\lenspotscalar,\rm st}=5, \lambda_{\lenspotscalar,\rm bl}=6\}$. The resulting \lenspotpix models and corresponding residuals are shown in Fig.~\ref{fig:model_different_lambdas}.

Low-regularization models lead to a slight over-fitting as seen from the reduced $\chi^2$ below unity. The reconstructed perturbations display higher frequency features, some being present in the input perturbations as in lens PS, while some others are artifacts as in lenses LS and HM. Increasing the regularization strength filters out the high-frequencies, leading to smoother variations in the \lenspotpix maps. For highest regularization strengths, more features are visible, which is translated in higher $\chi^2$ values. Moreover, the amplitude of the modeled perturbations decreases as the regularization strength increases, which is also expected since high-frequencies, which are suppressed, have the highest amplitudes. Overall, the main features of the perturbations are well-recovered for these three reasonable choices of hyper-parameters $\lambda_{\lenspotscalar,\rm st}$ and $\lambda_{\lenspotscalar,\rm bl}$. These results also confirm that our fiducial setting is close to be optimal since it provides a good compromise between over-fitting and correctly fitting the data.

\subsection{Smooth potential parameters \label{ssec:smooth_parameters}}

\begin{figure*}[!tbh]
    \centering
    \includegraphics[width=\linewidth]{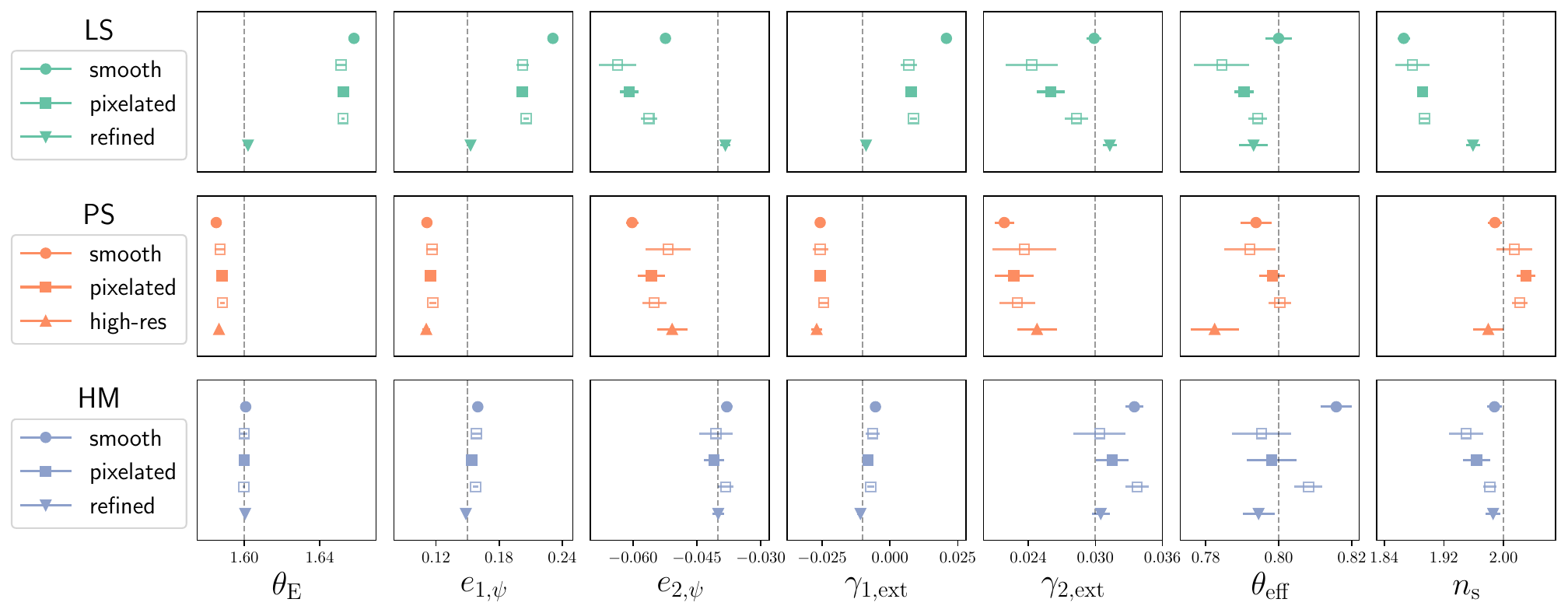}
    \caption{MAP values and estimated uncertainties for the smooth potential parameters $\modelparam_\lenspotscalar$ and a subset of the source parameters $\modelparam_\source$. From left to right: Einstein radius, complex ellipticities of the SIE, external shear components, source effective radius, and Sérsic index. The dashed lines indicate the input values. The ``smooth'' models correspond to models without including pixelated perturbations; ``pixelated'' models correspond to our baseline models shown in Fig.~\ref{fig:model_summary_pix_pot_real}; ``high-res'' (lens PS) is similar to the fiducial model but with a smaller \lenspotpix pixel size; ``refined'' corresponds to models where the pixelated perturbations have been replaced with an analytical profile (see text for more details). For ``pixelated'' models we also show model parameters obtained with different regularization strengths with empty square symbols, above (low strength) and below (high strength) the fiducial model (see Fig.~\ref{fig:model_different_lambdas}). For several parameters, error bars are smaller than the marker size.}
    \label{fig:smooth_params}
\end{figure*}

In the previous section, we have seen that the main features of the perturbations are overall well recovered, but the reconstructions are not perfect despite model residuals almost at the noise level. Therefore, we expect that some of the inaccuracies in the pixelated potential model are absorbed by the smooth analytical component of the lens potential, or vice versa.

We check this hypothesis by comparing different models of the smooth potential, with and without including potential perturbations. As discussed earlier in this section, parameter uncertainties are either computed using the fast approximation of the FIM (fully smooth models), or based on HMC sampling of the parameter space (models including pixelated perturbations). The MAP values and uncertainties of a subset of analytical parameters are shown in Fig.~\ref{fig:smooth_params}, and compared against the input values. We discuss here only lens potential parameters, but the conclusions are similar for the source parameters as well, that we present in Appendix~\ref{app:sec:all_smooth_parameters}.

We observe that fully smooth models display strong biases in almost all parameters as expected. Interestingly, models including a pixelated model in the potential, while having larger error bars, still lead to statistically significant biases. One particularly informative parameter is the inferred Einstein radius, as the value of $\theta_{\rm E}$ is slightly more accurate after including the potential perturbations, but a substantial bias still remains. We attribute those biases to a manifestation of the degeneracies that exist between the smooth component of the lens potential and the pixelated perturbations, where some adjustments of one component can compensate for the other, still leading to comparable model residuals \citep[see][who also observed reabsorption of residuals in the macro potential model]{Vegetti2014}. Additionally, we see that the regularization strengths of the pixelated potential marginally impact the results. Parameter uncertainties are increased with lower regularization strength, which is expected as the model has effectively more degrees of freedom that are not fully constrained.

\subsection{Analytically refined models \label{ssec:refined_models}}

To assess if the biases discussed above can be mitigated, we reduce the parameter space by replacing the pixelated component with analytical profiles. In a real-world scenario, where the underlying type of perturbations is unknown, this would correspond to imposing stronger priors on the model, motivated by the characteristic features observed in the pixelated model. This strategy is similar to previous studies based on the gravitational imaging technique \citep[e.g.][]{Vegetti2009,Vegetti2012}, where an analytical profile is optimized at the position of a tentative detection of a subhalo, in order to better characterize its properties (position, radial profile, mass).

Among the different systems modeled in this work, the most characteristic features that can be noticed in the pixelated models are for the LS and HM cases, with either a very localized decrease of the potential, or azimuthal periodicity centered on the lens galaxy, respectively. Therefore, for these two lenses, we start from the MAP solution obtained with our fiducial model and include a SIS profile (lens LS) or an multipole component (lens HM) instead of the pixelated component. We refer to these additional fully analytical models as ``refined'' in the remaining of the text.

The smooth potential parameters inferred from these refined models are compared to the smooth models in Fig.~\ref{fig:smooth_params}. We see that the significant biases observed with models with too few or too many degrees of freedom have been correctly mitigated. This result is not surprising as the refined model is now parametrized identically to the simulated data. Nevertheless, this allows us to confirm that even after optimizing a more complex and possibly inaccurate model, it does not prevent us from accessing the optimal global solution after correctly identifying the underlying type of perturbations. In Sect.~\ref{sec:model_properties} we use these refined models to retrieve the properties of the underlying perturbations.

\subsection{Higher resolution pixelated model \label{ssec:refined_models}}

Contrary to the localized subhalo and multipolar structures, the underlying perturbations for lens PS are described by a GRF, which does not have a specific analytical profile as it is a random realization. Instead, we use this system to test if using a higher resolution grid for the pixelated component \lenspotpix (i.e., more parameters) allows us to reduce the biases on smooth model parameters. The resulting model, named ``high-res'', is compared to the fully smooth and fiducial models in Fig.~\ref{fig:smooth_params}. While this model is insufficient to recover unbiased smooth potential parameters, we subsequently see in Sect.~\ref{sec:model_properties} that the recovered power-spectrum of the perturbations is closer to the input one.

Nevertheless, we note that recovering the smooth potential parameters in the presence of perturbations such as a subhalo population modeled via a GRF is achievable by imposing informed priors on the pixelated potential model. This has been shown in the recent work of \citet{Vernardos2022}, which extended the gravitational imaging technique using a covariance-based regularization of the pixelated potential model. The covariance matrix governing the regularization term can be specifically adapted to GRF-like perturbations, leading to an effective regularization of the solution if the assumption matches the underlying perturbations. We plan to implement the strategy presented in \citet{Vernardos2022} in the \herculens package, and leave for future works its comparison with the method presented here.

\begin{table*}[!tbh]
\caption{Recovered properties of potential perturbations, from the MAP \lenspotpix models shown on Fig.~\ref{fig:model_summary_pix_pot_real}. The quoted values for the subhalo mass is given in solar units and computed within the region with S/N(\lenspotpix) > 3 (see Fig.~\ref{fig:subhalo_position_mass} and text for more details). The input GRF parameters are given with uncertainties as they were fitted on the power-spectrum of the input perturbing field.}
\label{tab:param_estimations}
\renewcommand{\arraystretch}{1.8}
\centering
\small{
\begin{tabular}{lcclc}
Lens & Parameters & Input values & Model & Measured values \\
\hline
\hline
\multirow{4}*{LS} & \multirow{2}*{$\big(\theta_{x,\rm sub},\theta_{y,\rm sub}\big)$} & \multirow{2}*{$\big( 1.90, -0.40 \big)$} & pixelated & $\big( 1.94 \pm 0.12, -0.49 \pm 0.12 \big)$ \\
 & & & refined & $\big( 1.899 \pm 0.008, -0.408 \pm 0.005 \big)$ \\
\arrayrulecolor{gray}\cline{2-5}
 & \multirow{2}*{$\log_{10}M_{\rm sub}$} & \multirow{2}*{$8.91$} & pixelated & $8.71 \pm 0.05$ \\
 & & & refined & $8.90 \pm 0.01$ \\
 
\arrayrulecolor{black}\hline

\multirow{3}*{PS} & \multirow{3}*{$\big( \log_{10}\sigma^2_{\rm GRF}, \beta_{\rm GRF} \big)$} & \multirow{3}*{$\big( -3.11\pm0.15, 3.23\pm0.05 \big)$} & pixelated, ideal & $\big( -3.35\pm0.20, 3.40\pm0.15 \big)$ \\
 & & & pixelated, fiducial & $\big( -3.89\pm0.25, 4.15\pm0.19 \big)$ \\
 & & & pixelated, high-res & $\big( -3.49\pm0.15, 3.26\pm0.09\big)$ \\
 
\arrayrulecolor{black}\hline

\multirow{3}*{HM} & \multirow{2}*{$\big(a_m,\phi_m\big)$} & \multirow{2}*{$\big( 0.06, 82.50 \big)$} & pixelated & $\big( 0.024 \pm 0.002, 83.78 \pm 0.88 \big)$ \\
 & & & refined & $\big( 0.061 \pm 0.002, 83.15 \pm 0.39 \big)$ \\
 \arrayrulecolor{gray}\cline{2-5}
 & $m$ & $4$ & refined & $4.01 \pm 0.02$ \\
 
\arrayrulecolor{black}\hline
\end{tabular}
}
\end{table*}

\section{Constraints on the underlying perturbations \label{sec:model_properties}}

In the previous section we discussed in a qualitative manner the reconstructed perturbations. Here we seek to quantify the properties that can be recovered from these models, and discuss the robustness of those measurements as well as their applicability to real data sets. All the inferred quantities discussed in the next subsections are summarized in Table~\ref{tab:param_estimations}, and based on the fiducial models shown in Fig.~\ref{fig:model_summary_pix_pot_real}.

\subsection{Subhalo mass and position \label{ssec:analyse_lens_LS}}

We consider the two models of lens LS to quantify the properties of the underlying DM subhalo: our fiducial model including a pixelated component in the lens potential, and the refined model assuming the detected subhalo mass distribution follows a SIS profile.

\begin{figure*}[!tbh]
    \centering
    \includegraphics[width=0.7\linewidth]{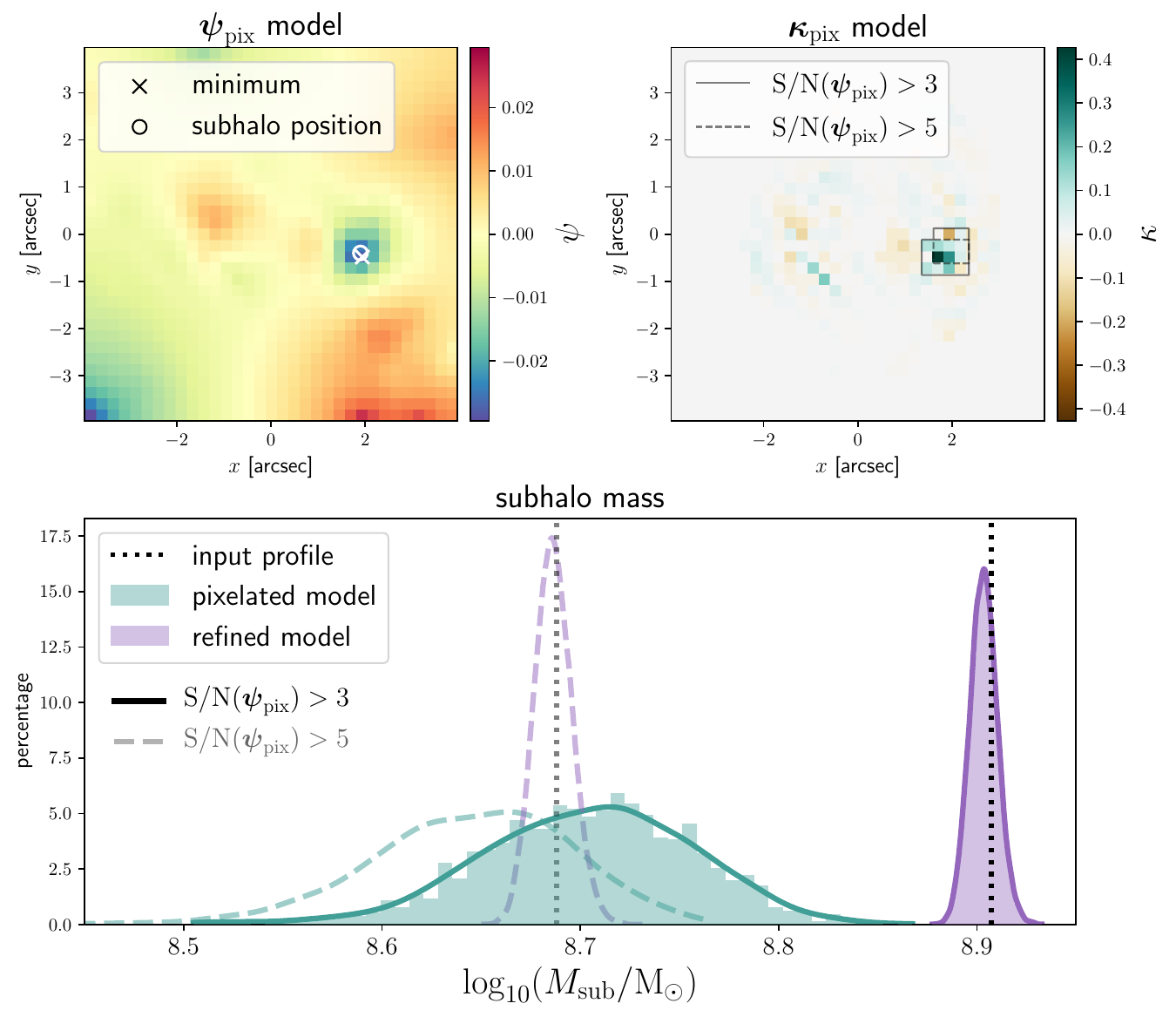}
    \caption{Characterization of the position and mass of the localized subhalo based on our model of lens LS. \textit{Top left}: \lenspotpix model, from which we assign the detected subhalo position to the minimum of the clear negative feature, in good agreement with the input subhalo position. \textit{Top right}: pixelated convergence model \convpix obtained from the \lenspotpix model. We indicate the fiducial pixelated region within which we compute the subhalo mass, defined as the pixels with S/N(\lenspotpix) > 3 (see also the bottom panel of Fig.~\ref{fig:model_summary_pix_pot_real}). A smaller region corresponding to S/N(\lenspotpix) > 5 is also shown, in dashed lines. \textit{Bottom}: posterior distributions of the subhalo mass, as estimated from the pixelated model, and from the refined model that replaces \lenspotpix with a SIS profile for the subhalo. The two pairs of posterior distributions corresponding to each pixelated region are shown in continuous (S/N(\lenspotpix) > 3) or dashed lines (S/N(\lenspotpix) > 5). For each pixelated region, the black dotted line indicates the mass computed from the input convergence profile of the subhalo.}
    \label{fig:subhalo_position_mass}
\end{figure*}

For the pixelated model we assign the position of the detected subhalo to the minimum of the pixelated potential \lenspotpix. We show the location of the pixel in the top left panel of Fig.~\ref{fig:subhalo_position_mass}. For related uncertainties, we compute the minimum of each HMC sample of \lenspotpix, but we note that the minimum remains in the same potential pixel, leading to error bars smaller than its size. We thus turn to a more conservative estimate of the uncertainty and simply set it to half the pixel size (i.e., $0\farcs12$). For the refined model, we take the optimized position of the SIS as the position of the subhalo, with uncertainties estimated from the FIM. The resulting positions and error bars are listed in the top row of Table~\ref{tab:param_estimations}.

The mass of the subhalo is more difficult to estimate from our pixelated model. Nevertheless, achieving this would be powerful, as it does not require the choice of a specific shape for the subhalo mass distribution. We start by computing the surface mass density (i.e., the convergence) corresponding to each potential pixel using Eq.~\ref{eq:convergence}. This results in the pixelated convergence model \convpix shown in the top right panel of Fig.~\ref{fig:subhalo_position_mass}. Next, we need to define a region in which to integrate \convpix pixels before converting the surface mass density to proper solar mass units using $\Sigma_{\rm crit}$. As discussed in Sect.~\ref{ssec:lens_LS}, input parameters of the subhalo correspond to a subhalo mass of about $10^9$ within the Einstein radius of $0\farcs07$. As this scale is much smaller than a pixel of our \lenspotpix model, we cannot rely on summing the convergence pixels.

We address this issue by considering a larger region within which the subhalo mass can be inferred for both the analytical and pixelated models. We select a high-significance region of the reconstruction that contains all pixels with S/N(\lenspotpix) > 3 (see bottom left panel of Fig.~\ref{fig:model_summary_pix_pot_real}). This region contains 15 convergence pixels, that we sum  and convert to proper units to estimate the subhalo mass. We repeat the same procedure for each sample of the joint posterior distribution ($\sim 1300$ samples) and find the distribution shown in the bottom panel of Fig.~\ref{fig:subhalo_position_mass}. To compare the inferred value with that of the input subhalo, we disctretize the input SIS profile by evaluating it on the same grid of pixels, and compute the mass as for the pixelated model. We note that the resulting ``input'' mass is lower than the one computed analytically within Einstein radius, because of the discretization of the SIS profile that diverges in the center.

We find that the inferred mean value of the subhalo is lower than the measured mass on the input perturbations. In addition, we find that the amplitude of the disagreement is significantly affected by the choice of the region in which we integrate the convergence. For instance, considering only \convpix pixels with S/N(\lenspotpix) > 5 instead of 3 (6 pixels instead of 15) leads in fact to a very nice agreement with the input value. This assumption corresponds to the dashed-line histograms in Fig.~\ref{fig:subhalo_position_mass}. The main reason of the better agreement is that this smaller region essentially excludes the few pixels with negative convergence, seen in brown color in Fig.~\ref{fig:subhalo_position_mass} (negative convergence pixels do have a physical meaning, as they indicate a local decrease of the lens mass relative to the smooth component). On one hand, this leads to a higher mass inferred from the model; on the other hand, the input value used as a reference is smaller, because it is computed within a smaller region. These two effects combined lead to an overall better agreement between the model and the input. We note that a similar behavior is observed when the \lenspotpix regularization strength is too low (see Sect.~\ref{ssec:effect_lambda}). However, in this case, the pixelated convergence map is very noisy due to over-fitting the imaging data and the region inside which the subhalo mass is measured cannot be reliably defined.

Measuring the mass directly from the pixelated model is therefore challenging and possibly depends on additional assumptions. Currently, a more robust approach is to infer the subhalo mass from our refined model, which is based on an analytical profile for the subhalo. After applying the same procedure, we obtain the resulting posterior distribution shown in purple in Fig.~\ref{fig:subhalo_position_mass}, which is in perfect agreement with the input. Again, this is not surprising, as the SIS profile reflects well the underlying shape of the subhalo. Nevertheless, these results showcase the requirement of stronger priors on the shape of the subhalo in order to infer unbiased properties. Additionally, for proper inference on real data, several works have demonstrated the need to carefully compare different choices of subhalo profiles \citep[e.g.][]{Sengul2021,Despali2022}.

\subsection{Statistics of the subhalo population}

\begin{figure}[!t]
    \centering
    \includegraphics[width=\linewidth]{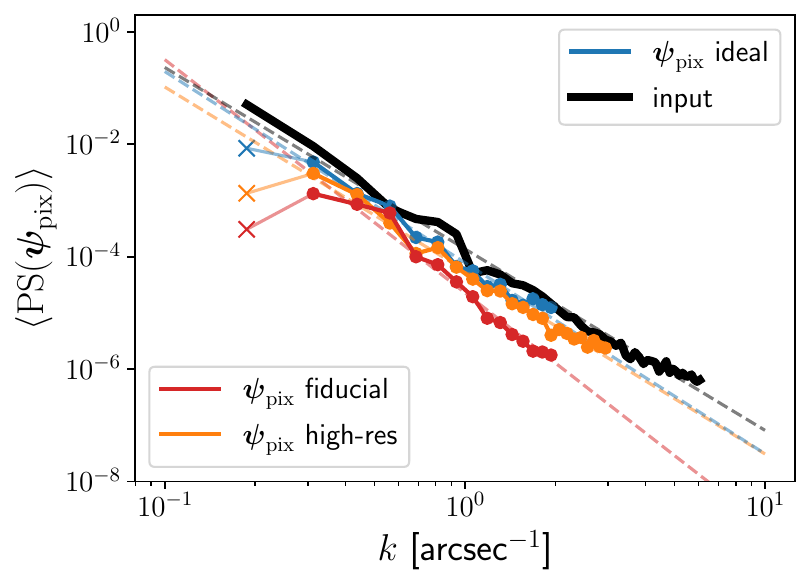}
    \caption{Azimuthally averaged power spectra of input and modeled (\lenspotpix) perturbations for lens PS (computed within the same mask as in Fig.~\ref{fig:model_summary_pix_pot_real}). Linear relations used to estimate the variance $\sigma^2_{\rm GRF}$ and slope $\beta_{\rm GRF}$ are overlayed (dashed lines), assuming the underlying perturbing field is a GRF (uncertainties are estimated from the least-squares fit). The crosses indicate the bin ($2\times10^{-1}$) that is ignored for the linear fit, as it corresponds to the size of the mask.}
    \label{fig:power_spectra}
\end{figure}

We analyze our pixelated reconstruction for lens PS following a Fourier power-spectrum analysis, motivated by the assumption of GRF perturbations (Eq.~\ref{eq:def_grf}). We show in Fig.~\ref{fig:power_spectra} the azimuthally averaged power-spectra from the three different pixelated models explored in this work: the ``ideal'' (Fig.~\ref{fig:model_summary_pix_pot_ideal}), the ``fiducial'' (Fig.~\ref{fig:model_summary_pix_pot_real}), and the ``high-res'' models (i.e., finer \lenspotpix pixels, twice the data pixels). We compute the power-spectra inside the region of Fig.~\ref{fig:model_summary_pix_pot_real}, in order to consider only features in the region of interest. We then compare the obtained power-spectra with those from the input perturbations by fitting linear relations in log-space and list the resulting best-fit values for $\sigma_{\rm GRF}^2$ and $\beta_{\rm GRF}$ in Table.~\ref{tab:param_estimations}. We note that the first bin is excluded from the linear fit because it corresponds to a wavenumber that translates to the roughly the size of the region used for computing the power-spectra, hence it is no informative. The quoted uncertainties are estimated from the least-square fit.

As expected, the model in the ideal case (i.e., with fixed smooth potential and source light) agrees very well with the input power-spectra, which translates in a good agreement for GRF parameters as well. Regarding our fiducial model, the amplitude is overall lower than the input, consistent with we what we discussed in Sect.~\ref{sec:modeling_results}. At intermediate wavenumbers, the recovered power-spectrum is close to the input one, however it is strongly attenuated at large wavenumbers. This leads to an overall steeper slope, and translates to a $\sim$ $5\sigma$ difference in $\beta_{\rm GRF}$ with respect to the input. This attenuation of small spatial scales is fully mitigated by modeling the perturbation on a higher-resolution grid, that allows us to better model small scale features. Indeed, the power-spectrum of the high-res model exhibits an excellent agreement with the input for all wavenumbers $k > 4\cdot10^{-1}$ arcsec$^{-1}$. The inferred slope $\beta_{\rm GRF}$ is within $1\sigma$ with respect to the input value (see Table~\ref{tab:param_estimations}).

Overall, our pixelated method correctly retrieves the locations of the main perturbations that mimic a subhalo population, and allows us to obtain a first-order estimate of its statistics. Our results suggest that using a higher-resolution grid for the pixelated component allows to better recover the full power-spectrum. However, the precise characterization of the power-spectrum of the perturbations under the assumption of GRF is a challenging task that requires additional priors in the model \citep[e.g.,][]{Bayer2018,Vernardos2022}.

\subsection{Properties of multipolar structures}

Based on our models of lens HM, we can recover the underlying octupole using two different methods: fitting an octupole directly on the \lenspotpix model, or using the refined model that includes explicitly an octupole profile in addition to the other components of the lens potential.

\begin{figure}[!t]
    \centering
    \includegraphics[width=0.8\linewidth]{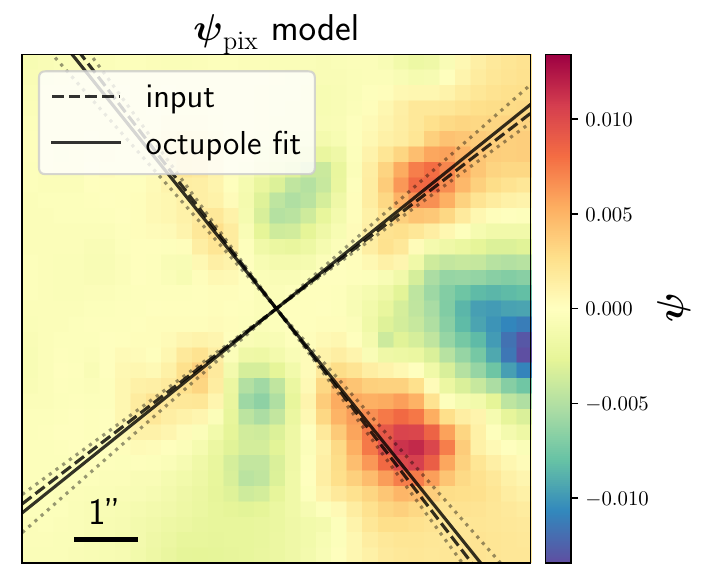}
    \caption{Measurement of the multipole orientation from the \lenspotpix model of lens HM. An analytical octupole profile (plus a constant offset) is fitted directly on the reconstructed perturbations, and compared to the input one. Each pair of lines shows two of the main octupole axes, to ease the visualization. The gray dotted lines are corresponds to $\pm3\sigma$ fits, where $\sigma$ is estimated from the FIM.}
    \label{app:fig:multipole_orientation}
\end{figure}

We perform the octupole fitting (i.e., we fix $m=4$) on the \lenspotpix model via gradient descent with three free parameters, namely the amplitude, orientation and an additional constant offset (remember that this offset in the potential is not constrained by the data). After converging to the MAP solution, parameter uncertainties are estimated from the FIM. The recovered octupole orientation, reported in Table~\ref{tab:param_estimations}, agrees extremely well with the input. However, the recovered amplitude is lower, as expected from the pixelated reconstructions and discussed in Sect.~\ref{sec:modeling_results}.

With the refined model it is possible to measure the constraints on the multipole order $m$, in addition to the orientation and amplitude. The MAP values of the multipole component obtained with this refined model are reported in Table~\ref{tab:param_estimations}, which all agree very well with the input. The parameter $m$ is expected be more challenging to optimize, as it has nonlinear effects on the profile shape. We tested the robustness of the optimization to different initial values $m$, and found that initializing $m$ to a value of 3 or higher leads to the correct MAP value, but setting it closer to 2 drives the model toward a quadrupole ($m=2$), which is degenerate with the shear and ellipticity of the smooth potential.

\section{Computation time}

\herculens uses \jax to exactly differentiate the loss function and significantly decrease runtime (Sect.~\ref{ssec:autodiff}). The entire analysis of this work, including parameters optimization and sampling, was performed on a personal computer. We give average timings of the main modeling steps for a personal computer\footnote{We note that most of the timings quoted here also include an overhead time of about 2 to 4 seconds, due to \jax ``just-in-time'' compilation feature.}:
\begin{itemize}
    \item Optimizing the smooth analytical models (12 parameters), takes one minute for a multistart gradient descent with 30 starts (i.e., $\sim 2$ seconds for a single gradient descent).
    \item Computing the regularization weights (Eq.~\ref{eq:wavelet_regul_pot}) takes about 20 seconds. This step can be accelerated at least by a factor of two, by precomputing some of the operators involved in Eq.~\ref{eq:deta_d_delta_psi_func}; we leave this for future improvements.
    \item Optimizing the idealized models of Fig.~\ref{fig:model_summary_pix_pot_ideal} in which only the pixelated potential component is optimized (1089 parameters) takes 30 seconds. This is for $10^3$ iterations, which is sufficient to reach convergence.
    \item Compared to these idealized models, the run time is virtually identical for optimizing the full models of Fig.~\ref{fig:model_summary_pix_pot_real}, due to the marginal increase in the number parameters (1101 parameters).
    \item Computing the FIM and its inverse typically takes 20 seconds.
    \item Performing HMC sampling for a total of 500 samples takes about 1.1 hours for a single chain.
\end{itemize}
These numbers can be extended to the modeling of a typical HST observation of a strong lens. For instance, modeling the lens SDSS\,J$1630+4520$ \citep[from the SLACS sample,][]{Bolton2006}, with a smooth lens potential and a pixelated source regularized with wavelets \citep[as in][]{Galan2021} showed that convergence to the MAP takes about 1.5 minutes for a single gradient descent (still on a personal computer). This includes preoptimization steps with a smooth model for source whose complexity is progressively improved with a pixelated source. Fitting the lens light with analytical profiles would only lead to a marginal increase of the total run time ($\sim 10$ seconds). Next, modeling deviations to the smooth lens potential assuming the initial model fits reasonably well the data would require about 30 seconds for a single gradient descent, similar to the models presented in this work. Finally, sampling the full parameter space using HMC would take from one to two hours for $\sim\mathcal{O}(10^3)$ samples, which we expect to be sufficient to ensure well-sampled posterior distributions.

While the timings quoted above demonstrate that our code can be conveniently run on a single CPU, they do not reveal the full potential of the method. \herculens is fully based on \jax so it supports large scale parallelization and GPU capabilities, which can lead to dramatic improvements in terms of computation time \citep[see e.g.,][]{Gu2022}. This will be exploited in future analyses of real data sets.

\section{Summary and conclusion \label{sec:discussion_conclusion}}

In this work we develop and apply a novel lens modeling method that is able to recover perturbations to a mostly smooth lens potential with minimal assumptions about their nature. This is made possible by modeling the perturbations on a grid of pixels regularized using a well-established multiscale technique based on sparsity and wavelets. This grid of pixels is superimposed on other analytical profiles for the joint modeling of the full potential and the source light. We show that merging the two main state-of-the-art modeling paradigms (analytical and pixelated) is possible within the framework of differentiable programming. This enables us to seamlessly optimize lens models with over one thousand individual parameters and obtain their uncertainties, either via Fisher information analysis or gradient-informed HMC sampling.

We summarize the key results of this work as follows:
\begin{itemize}
    \item We extend our previous work in  \citet{Joseph2019} and \citet{Galan2021}, by introducing a pixelated mass component in addition to the smooth lens potential, and recover three different types of perturbations: a localized DM subhalo, a population of such subhalos, and high-order moments in the lens potential.
    
    \item Sparse regularization is usually performed via iterative algorithms to converge to the solution, which makes it challenging to incorporate within standard lens modeling codes. In this work we demonstrate that the solution can also be obtained via gradient descent, capitalizing on the access to derivatives of the loss function with automatic differentiation.
    
    \item Differentiable programming enables the simultaneous optimization of analytical and pixelated components. One benefit is that the perturbative approach of \citet{Koopmans2005} for pixelated perturbations to the lens potential is no longer warranted, although we can still use it to estimate regularization weights. This is because the full inverse problem can now be solved from the explicit superposition of smooth analytical and pixelated components, jointly optimized via gradient descent.
    
    \item For each type of perturbation explored in this work, the main features are correctly recovered by the pixelated potential component. In particular, the signature of a localized subhalo and octupolar structures are accurately captured. The subhalo population, simulated here as a GRF, clearly represents a more challenging situation, although the main over- and under-density regions can still be recovered.
    
    \item We test for a model-independent recovery of the DM subhalo mass, directly from the reconstructed pixelated potential model. We find that the resulting measurement of the mass is sensitive to the region where the surface density is integrated, leading to either an under-estimation of the mass, or a value in agreement with the true subhalo mass. Nevertheless, switching to an analytical profile for the subhalo, as is standard practice, allows to robustly infer its mass.
    
    \item The statistical properties of the subhalo population can be recovered via the power spectrum of the pixelated model. The underlying GRF parameters, which effectively act as parameters of the subhalo mass function, are recovered with our higher resolution model. Nevertheless, in a real-world scenario, we advocate for a comparison of different model variants, typically with different grid resolutions, and a cross-checking with other methods relying on more informative priors \citep[e.g., as in][]{Vernardos2022} to improve the robustness of the inference.
    
    \item High-order moments in the lens potential (here as an octupole) are remarkably well recovered, despite being small in amplitude compared to lower-order moments such as quadrupoles (e.g., external shear). While the amplitude is biased low in the pixelated reconstruction, the octupole orientation is accurately measured, either from the pixelated model, or using a refined model including a multipole component in the lens potential explicitly.
\end{itemize}

Our method is readily applicable to real HST observations of EELs, such as the systems presented in \citet{OldhamAuger2018}, as the source surface brightness is smooth. This assumption of smoothness, although well motivated by real observations, is arguably the strongest assumption of this work. Indeed, there are many situations in which the source galaxy is more complex, featuring for instance spiral arms and localized clumpy star forming regions. This requires the joint modeling of deviations from smoothness both in the source and the lens potential, which is challenging due to degeneracies between those two components, as some of the lensing features may be equally well explained by underlying features in the source or in the potential.

Nevertheless, the modeling methods presented in this work and implemented in \herculens enable the joint modeling of more complex sources on a grid of pixels in the source plane, which we will apply in future analyses. Moreover, the flexible framework provided by differentiable programming allows one to implement a large fraction (if not all) of the modeling methods explored in the literature so far. The careful comparison between their different assumptions is absolutely critical to mitigate degeneracies and robustly infer key physical quantities, including constraints on the subhalo mass function and the mass of DM particles. Recently, there have been a few works along these lines, including the reanalysis of a subhalo detection based on different assumptions such as the shape of the subhalo mass profile and its redshift \citep{Sengul2021}, or the thorough comparison of different modeling codes on the same data set \citep{Shajib2022_TDC_9}.

In this work we limit ourselves to three categories of perturbations in the lens potential, however there exist others. Another departure from the simple elliptical profiles is the presence of a disk component in the lens galaxy. In particular, such a disk structure can have observational effects that are similar to flux ratio anomalies in multiply imaged lensed quasars. These anomalies are usually considered as a tracer for DM subhalos, hence unique probes of the subhalo mass function. In a series of papers based on cosmological simulations and real observations, \citet[and references therein]{Hsueh2018} showed that not taking into account these baryonic effects on the lens potential can bias the inferred constraints on DM properties. The detection and modeling of a disk component in the lens potential is therefore a situation where our pixelated reconstruction method could be successful.

\herculens is built on top of differentiable programming libraries that readily enable massive parallelization and GPU acceleration. This important aspect has been extensively discussed in \citet{Gu2022}, where authors demonstrated the efficiency of gradient-informed methods for modeling large samples of strong lenses. Their results remain relevant and are applicable in the context of our work as well. This is a promising path towards the modeling of the tens of thousands strong lenses that will be discovered in the era of Vera Rubin and Euclid survey telescopes \citep{Collett2015}. Moreover, highly flexible yet computationally efficient models will be essential to handle the even higher resolution data sets soon delivered by the James Webb Space Telescope and future extremely large telescopes.

\begin{acknowledgements}
We thank the anonymous referee for helpful comments that improved this manuscript. We thank Luca Biggio, Martin Millon, Eric Paic and Simona Vegetti for useful feedback and discussion on the present work. We thank Simon Birrer for useful discussion, and for making the modeling software package \lenstro open-source, which made the development of \herculens much easier. This programme is supported by the Swiss National Science Foundation (SNSF) and by the European Research Council (ERC) under the European Union’s Horizon 2020 research and innovation programme (COSMICLENS: grant agreement No 787886). 
GV has received funding from the European Union’s Horizon 2020 research and innovation programme under the Marie Sklodovska-Curie grant agreement No 897124.
This research has also made use of \textsc{SciPy} \citep{Virtanen2020scipy}, \textsc{NumPy} \citep{Oliphant2006numpy,VanDerWalt2011numpy}, \textsc{Matplotlib} \citep{Hunter2007matplotlib}, \textsc{Astropy} \citep{astropy2013,astropy2018}, \textsc{Jupyter} \citep{Kluyver2016jupyter} and \textsc{GetDist} \citep{Lewis2019getdist}.
\end{acknowledgements}


\bibliographystyle{aa}
\bibliography{biblio}


\begin{appendix}

\section{Multiscale regularization weights \label{app:sec:regularization_weights}}

In this section we give details about the estimation of the weights in Eq.~\ref{eq:wavelet_regul_pot} used to scale the regularization strength of the pixelated potential \lenspotpix. Let us assume that we have in hand a model fitted to the data without \lenspotpix, such that we can assume that a sufficiently good model of the source \source and the smooth lens potential are known.

The minimization problem corresponding to the full lens potential, parametrized as $\lenspot = \tilde{\lenspot}(\modelparam_\lenspot) + \lenspotpix$ can be written as
\begin{align}
\small
    \label{eq:argmin_pb_lenspot}
    \nonumber
    \argmin_{\modelparam_{\lenspot},\lenspotpix}\ &\frac12\, \Big[\,\data - \convop\lensingop\source\,\Big]^\top \datacovmatrix^{-1}\ \Big[\,\data - \convop\lensingop\source\,\Big] \\
    &+ \lambda_{\lenspot}\,\normone{ \weights_{\lenspotscalar} \circ \waveletop^\top\, \lenspotpix} \ ,
\end{align}
where \lensingop depends on the smooth potential parameters $\modelparam_{\lenspot}$ and the pixelated component \lenspotpix. The regularization term is identical to Eq.~\ref{eq:wavelet_regul_pot}.

In contrast to the source reconstruction problem, namely where the variable of interest is \source, the model we optimize here does not depend linearly on $\modelparam_{\lenspot}$ and $\lenspotpix$. Consequently, the gradient of the data-fidelity term in Eq.~\ref{eq:argmin_pb_lenspot} does not have a simple closed-form expression, which prevents us from propagating the noise from the data to the lens potential as easily as to the source plane.

We address this issue by considering the perturbative approach of \cite{Koopmans2005} as a means to relate the data space to the potential space. We argue that we can use the weights obtained with this model for the minimization problem above.

Based on a Taylor expansion of the lens equation with respect to small deviations $\delta\lenspot$ to an underlying smooth potential $\tilde\lenspot(\modelparam_{\lenspot})$, \cite{Koopmans2005} established the following linear relation
\begin{align}
    \label{eq:deta_d_delta_psi_op}
    \delta\data &= \underbrace{-\convop\,\op{D}_{s}\!\left(\source\right)\op{D}_{\delta\psi}}_{\mbox{\op{D}}}\,\delta\lenspot \ ,
\end{align}
which corresponds to Eq.~\ref{eq:deta_d_delta_psi_func} rewritten with linear operators. The term $\delta\data \equiv \data - \tilde\model$ represents residuals between the data and a model $\tilde\model = \convop\lensingopsmooth\source$ without perturbations to the smooth potential (i.e., $\delta\lenspot = 0$). The operator $\op{D}_{s}$ contains spatial derivatives of the source light model with respect to source plane coordinates mapped from the data grid via \lensingopsmooth. The operator $\op{D}_{\delta\psi}$ combines bilinear interpolation and finite difference coefficients to compute the spatial derivatives of $\delta\lenspot$ on the data grid. We refer the reader to \citet{Koopmans2005} for more details about the exact structure of these operators, that we implement as matrices.

From the linear relation of Eq.~\ref{eq:deta_d_delta_psi_op}, we can formulate yet a new minimization problem for $\delta\lenspot$ as
\begin{align}
\small
\label{eq:argmin_pb_delta_lenspot}
\nonumber
    \argmin_{\delta\lenspot}\ &\frac12\, \Big[\,\delta\data - \op{D}\delta\lenspot\Big]^\top \covmatrix_{\delta\data}^{-1}\ \Big[\,\delta\data - \op{D}\delta\lenspot\Big] \\
    &+ \lambda_{\delta\lenspot}\,\normone{ \weights_{\delta\lenspotscalar} \circ \waveletop^\top\, \delta\lenspot} \ ,
\end{align}
where $\covmatrix_{\delta\data}$ is the covariance matrix associated to $\delta\data$. From the definition of model residuals, we have
\begin{align}
    \label{eq:delta_d_cov}
    \nonumber
    \covmatrix_{\delta \data} &\equiv {\rm cov}\big(\delta \data, \delta \data\big) \\
    \nonumber
    &= {\rm cov}\big(\data - \tilde{\model}, \data - \tilde{\model}\big) \\
    \nonumber
    &= \datacovmatrix + \covmatrix_{\tilde{\model}} - 2\,{\rm cov}\big(\data, \tilde{\model}\big) \\
    &\approx \datacovmatrix \ .
\end{align}
The cross-covariance matrix ${\rm cov}\big(\data, \tilde{\model}\big)$ is zero, since the data and model $\tilde{\model}$ are uncorrelated (nature does not correlate with our model). Additionally, the covariance term $\covmatrix_{\tilde{\model}}$ is much smaller than $\datacovmatrix$ in the case of smooth analytical profiles for the potential and source (we checked this numerically). This is because the fully smooth model $\tilde{\model}$ is strongly limited by the shape of the analytical profiles described by a small number of degrees of freedom, constrained by a much larger number of data pixels. Although not explicitly specified for conciseness, the elements of \datacovmatrix\ depend on the model $\tilde\model$ to estimate the shot noise (Poisson noise) from the lensed features.

We now need to establish which operators are necessary to propagate the noise from $\delta\data$ space to wavelet coefficients of $\delta\lenspot$, in order to compute the coefficients of the matrix $\weights_{\delta\lenspot}$. We do so by considering a gradient descent update based on the data-fidelity term of Eq.~\ref{eq:argmin_pb_delta_lenspot} \citep[e.g.,][]{Lanusse2016}, which reads
\begin{align}
    \label{eq:gradient_step_delta_lenspot}
    \delta\lenspot^{(n+1)} = \delta\lenspot^{(n)} + \overbrace{\op{D}^\top\covmatrix_{\delta\data}^{-1} \left( \delta\data - \op{D}\delta\lenspot^{(n)} \right)}^{\text{negative data-fidelity\ gradient}} \ .
\end{align}
The above gradient translates model residuals (the term in parenthesis) into the coefficients of $\delta\lenspot$. Therefore, we see that the propagation of a given noise realization $\delta\data_N$ to the potential pixels is $\op{D}^\top\covmatrix_{\delta\data}^{-1}\delta\data_N$. We simply apply the wavelet transform operator to obtain wavelet coefficients instead, which leads to $\waveletop^\top\op{D}^\top\covmatrix_{\delta\data}^{-1}\delta\data_N$. Using Monte-Carlo simulations of $\delta\data_N$, we can compute the standard deviation $\sigma_{\delta\lenspot}^{(n,m,j)}$ for each potential pixel $(n,m)$ in each wavelet scale $j$, and populate the matrix $\weights_{\delta\lenspot}$ accordingly.

As a final step, we return to the original problem described by Eq.~\ref{eq:argmin_pb_lenspot}. In this problem, the pixelated potential parameters are \lenspotpix, instead of $\delta\lenspot$. First, let us assume that both \lenspotpix and $\delta\lenspot$ are defined on the same grid of potential pixels. Second, \lenspotpix is used in combination with a smooth lens potential component $\tilde{\lenspot}(\modelparam_\lenspotscalar)$, such that it captures deviations to the smooth potential in a similar way as $\delta\lenspot$. Third, from Eq.~\ref{eq:delta_d_cov}, we see that the noise in the data (characterized by \datacovmatrix) is a good approximation of the noise in the model residuals $\delta\data$ (characterized by $\covmatrix_{\delta\data}$). Therefore, the standard deviation of the noise in potential space, at the location of potential pixels and in each wavelet scale, can be approximated by the values $\sigma_{\delta\lenspot}^{(i,j,k)}$ computed above, namely $\weights_{\lenspotscalar} \approx \weights_{\delta\lenspotscalar}$.

In summary, we have used the linear expression of Eq.~\ref{eq:deta_d_delta_psi_op} as a way to set the weights of our regularization term (Eq.~\ref{eq:argmin_pb_lenspot}) based on the noise levels from the data. In this work, we use two different wavelet transforms, namely the starlet and Battle-Lemarié wavelet transforms. Thus we compute the corresponding weights following the above procedure to obtain $\weights_{\lenspotscalar,\rm st}$ and $\weights_{\lenspotscalar,\rm bl}$ respectively (only the wavelet operator $\waveletop^\top$ is different). We show in Fig.~\ref{fig:noise_wavelet_pixpot} the corresponding weights for each scale of the starlet transform.

\section{Analytical profiles \label{app:sec:analytical_models}}

The smooth models used throughout this work are based on analytical profiles, which we define in detail here. Elliptical profiles are described with coordinates $(\theta_1,\theta_2)$, obtained by rotating the original coordinates $\vec{\theta}\equiv(\theta_x,\theta_y)$ along the major axis of the ellipse with position angle $\phi$.

To describe the smooth component of the lens potential, we use the singular isothermal ellipsoid (SIE) profile. This profile is originally defined in surface mass density (convergence) but also has analytical formulae for the potential and deflection angles. The SIE potential is given by \citep[e.g.,][]{Keeton2001}
\begin{align}
\nonumber
    \psi_{\rm SIE}(\theta_1,\theta_2) &= \frac{\theta_{\rm E} \sqrt{q_{\rm m}}~\theta_1}{\sqrt{1-q_{\rm m}^2}}\, {\rm arctan} \left( \frac{\sqrt{1-q_{\rm m}^2}~\theta_1}{\sqrt{q_{\rm m}^2\theta_1^2+\theta_2^2}} \right) \\
    &+ \frac{\theta_{\rm E} \sqrt{q_{\rm m}} ~\theta_2}{\sqrt{1-q_{\rm m}^2}}\, {\rm arctanh} \left( \frac{\sqrt{1-q_{\rm m}^2}~\theta_2}{\sqrt{q_{\rm m}^2\theta_1^2+\theta_2^2}} \right) \ ,
\end{align}
where $\theta_{\rm E}$ is the Einstein radius, $q_{\rm m}$ is the axis ratio of the elliptical profile, and the corresponding position angle is $\phi_{\rm m}$. In practice, we do not optimize the axis ratio and position angle, but rather ellipticity components $\{e_{1,\psi}, e_{2,\psi}\}$ to prevent sampling issues with the periodicity of $\phi_{\rm m}$ (particularly for small values of $q_{\rm m}$). These components are defined as
\begin{align}
    \label{app:eq:def_ellipt}
    \begin{cases}
        e_{1,\psi} &= \frac{1-q_{\rm m}}{1+q_{\rm m}}\cos\left( 2\phi_{\rm m} \right) \\
        e_{2,\psi} &= \frac{1-q_{\rm m}}{1+q_{\rm m}}\sin\left( 2\phi_{\rm m} \right) \ .
    \end{cases}
\end{align}
The singular isothermal sphere (SIS) is a particular case of an SIE with no ellipticity ($q_{\rm m}=1$).

The influence of other galaxies in the vicinity of the main deflector---whose lensing effects remain linear with respect to the lensed source---is modeled as a uniform external shear. It has amplitude $\gamma_{\rm ext}$ and orientation $\phi_{\rm ext}$. The corresponding lensing potential in polar coordinates $(r,\phi)$ is
\begin{align}
    \label{eq:def_shear}
    \psi_{\rm ext}(r, \phi) = \frac12\,\gamma_{\rm ext}\,r^2\cos\left(2\phi - 2\phi_{\rm ext}\right) \ .
\end{align}
The corresponding shear ellipticity parameters that we used for optimization are defined as
\begin{align}
    \label{app:eq:def_ellipt}
    \begin{cases}
        \gamma_{1, \rm ext} &= \gamma_{\rm ext}\cos\left( 2\phi_{\rm ext} \right) \\
        \gamma_{2, \rm ext} &= \gamma_{\rm ext}\sin\left( 2\phi_{\rm ext} \right) \ .
    \end{cases}
\end{align}

The elliptical Sérsic profile, suitable for modeling smooth galaxy light distributions, is defined as \citep{Sersic1963}
\begin{align}
\label{eq:def_sersic_ellipse}
    I_{\rm Sersic}\left(\theta_1, \theta_2\right) = I_{\rm eff}\,\mathrm{exp}\!\left[ -b_n \left( \frac{\sqrt{\theta_1^2 + \theta_2^2/q_{\rm l}^2}}{\theta_{\rm eff}}\right)^{1/n_{\rm s}} + b_n \right] , 
\end{align}
where $I_{\rm eff}$ is the amplitude of the profile at the effective radius $\theta_{\rm eff}$, and $n_{\rm s}$ is the Sérsic index which defines the slope of the profile. The axis ratio is $q_{\rm l}$, and the corresponding position angle is $\phi_{\rm l}$. Ellipticities are defined as for the potential (see Eq.~\ref{app:eq:def_ellipt}). The term $b_n$ is not a free parameter and is computed such that $\theta_{\rm eff}$ is always equal to the half-light radius of the profile.

\section{Simulated HST observations \label{app:sec:sim_settings}}

In Table~\ref{app:tab:sim_params} we summarize all instrumental settings and specific model assumptions that were used to generate HST observations of EELs systems. We used the simulation software package \molet \citep{Vernardos2021} to perform high-resolution ray-tracing and to add instrumental effects (instrumental noise and PSF convolutions).

\begin{table}[!h]
\caption{Instrument settings and model assumptions used for generating HST/WFC3/F160W mock observations of EELs. The coordinates are oriented such as north is up and east is right. The orientation angle is east-of-north.} \label{app:tab:sim_params}
\renewcommand{\arraystretch}{1.2}
\centering
{\small
\begin{tabular}{lc}
\hline
\hline
\textbf{Observation} & \\
Pixel size [arcsec] & 0.08 \\ 
Exposure time [s] & $4\times2400 = 9600$ \\ 
Zero-point [mag] & 25.9463 \\ 
Sky brightness [mag/arcsec$^2$] & 22 \\ 
Read noise [e$^-$] & 4 \\ 
PSF (gaussian FWHM) [arcsec] & $0.3$ \\
Noise & readout + shot noise \\
\hline
\hline
\textbf{Deflector, $\zd=0.3$} & \\
\ \ \ Singular isothermal ellipsoid (SIE) & \\
\ \ \ \ \ \ Einstein radius, $\theta_{\rm E}$ [arcsec] & 1.6 \\
\ \ \ \ \ \ Ellipticity, $q_{\rm m}$ & 0.73 \\
\ \ \ \ \ \ Orientation, $\phi_{\rm m}$ & 82.5 \\
\ \ \ \ \ \ Position, ($x$, $y$) [arcsec] & $(0, 0)$ \\
\hline
\ \ \ External shear & \\
\ \ \ \ \ \ Strength, $\gamma_{\rm ext}$ & 0.032 \\
\ \ \ \ \ \ Orientation, $\phi_{\rm ext}$ & 144.2 \\
\hline
\ \ \ Sérsic (light) & \\
\ \ \ \ \ \ Ellipticity, $q_{\rm l}$ & $0.73\ (=q_{\rm m})$ \\
\ \ \ \ \ \ Orientation, $\phi_{\rm l}$ & $82.5\ (=\phi_{\rm m})$ \\
\ \ \ \ \ \ Half-light radius, $\theta_{\rm eff}$ & 2 \\
\ \ \ \ \ \ Sérsic index, $n_{\rm Sersic}$ & 2 \\
\ \ \ \ \ \ Position, ($x$, $y$) [arcsec] & $(0, 0)$ \\
\ \ \ \ \ \ Magnitude & 19 \\
\hline\hline
\textbf{Perturbations to the potential} & \\
\ \ \ \textbf{LS}: Localized subhalo (SIS) & \\
\ \ \ \ \ \ Einstein radius, $\theta_{\rm E,halo}$ [arcsec] & 0.07 \\
\ \ \ \ \ \ Position, ($x$, $y$) [arcsec] & $(1.9, -0.4)$ \\
\hline
\ \ \ \textbf{PS}: Population of subhalos (GRF) & \\
\ \ \ \ \ \ Variance, $\sigma^2_{\rm GRF}$ & $10^{-3}$ \\
\ \ \ \ \ \ Slope, $\beta_{\rm GRF}$ & $4$ \\
\hline
\ \ \ \textbf{HM}: High-order moments (octupole) & \\
\ \ \ \ \ \ Strength, $a_4$ & $0.06$ \\
\ \ \ \ \ \ Orientation, $\phi_{\rm 4}$ & $82.5=\phi_{\rm m}$ \\
\hline\hline
\textbf{Source, $\zs=0.7$} & \\
\ \ \ Sérsic & \\
\ \ \ \ \ \ Ellipticity, $q_{\rm l}$ & 0.82 \\
\ \ \ \ \ \ Orientation, $\phi_{\rm l}$ & 170.8 \\
\ \ \ \ \ \ Half-light radius, $\theta_{\rm eff}$ [arcsec] & 0.8 \\
\ \ \ \ \ \ Sérsic index, $n_{\rm Sersic}$ & 2 \\
\ \ \ \ \ \ Position ($x$, $y$) [arcsec] & $(0.4, 0.15)$ \\
\ \ \ \ \ \ Magnitude & 21 \\
\hline
\end{tabular}
}
\end{table}

\section{Choice of pixel size for \lenspotpix \label{app:sec:choice_pixel_scale}}

The pixelated model \lenspotpix used to capture deviations from the smooth lens potential requires a choice of pixel size. In \herculens, this can be set to any multiplicative factor (not necessarily an integer) of the data pixel size. In this work we use a pixel scale factor of 3 as our fiducial model, meaning the grid on which the \lenspotpix component is defined has a pixel size of $3\times0\farcs08 = 0\farcs24$. This resolution is sufficient to accurately model the simulated data set as well as to characterize the reconstructed features in the lens potential.

The multiscale aspect of the regularization strategy detailed in Sect.~\ref{ssec:wavelet_reg} enables the independent treatment of different spatial scales in the lens potential. This avoids the need to choose a specific pixel size, as long as it is small compared to the lensed features (otherwise it would lead to a poor fit to the data). Let us take the example of a pixel size much smaller than that of the data; if there are features below a given spatial scale that are not supported by the data (typically below the smallest detectable deflection angle), the multiscale wavelet decomposition combined with sparsity constraints will suppress all wavelet coefficients below this scale. Therefore, these features should not be visible in the reconstruction. This means that in comparing two models that fit the data equally well, we should select the one with the smaller pixel size. In practice, however, and from the perspective of parameter optimization, a smaller pixel size translates into a larger number of parameters, for which the regularization becomes even more crucial to balance the lack of constraints provided by the data alone. Moreover, the principle of Occam's razor, at the basis of Bayesian approaches, advocates for fewer parameters if it does not bring significant improvements to the fit.

\begin{figure*}[htbp!]
  \begin{adjustbox}{addcode={\begin{minipage}{\width}}{\caption{Best-fit models for different choices of pixel size of the \lenspotpix component, corresponding to 4, 3, 2, and 1.5 times the data pixel size. The modeling assumptions are otherwise identical to the results presented in Fig.~\ref{fig:model_summary_pix_pot_real} (the second row is exactly the same). From left to right, the columns show: model residuals with reduced $\chi^2$ value, best-fit \lenspotpix model, followed by each scale of the starlet transform of \lenspotpix. Each of these scales are images of the same size as \lenspotpix, above which we indicate the scale index $j$ and the characteristic scale $\tilde{\beta}_j$ (in arcsec) of the features. We see that a pixel scale of 3 is sufficient to model the full dynamical range of features supported by the data. Finer pixels are not required, as the smaller starlet scales reveal mainly low-significance artifacts over the level of individual pixels.
      }\end{minipage}},rotate=-90,center}
    \includegraphics[scale=0.55]{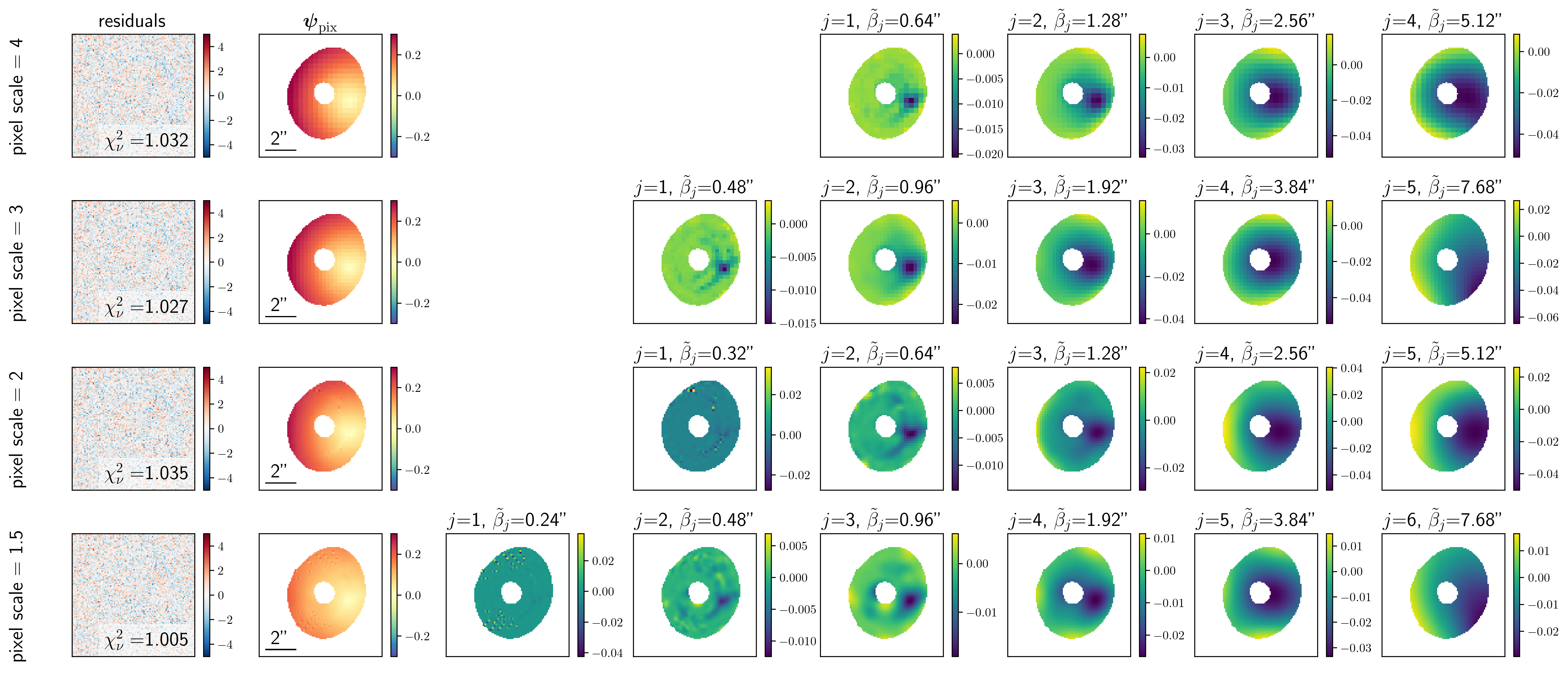}
    \label{app:fig:psipix_pixel_scale}
  \end{adjustbox}
\end{figure*}

We investigate the impact of the pixelated model resolution further by modeling lens LS (see Fig.~\ref{fig:data_summary}) using four different pixel sizes for \lenspotpix: 4, 3, 2 and 1.5 times the data pixel size (leading to 625, 1089, 2500 and 4489 parameters, respectively). For this simple exercise, we fix all other parameters to their input values. The resulting model residuals, best-fit \lenspotpix solution, and the full starlet decomposition of the solution are shown in Fig.~\ref{app:fig:psipix_pixel_scale}. For each of the starlet scales $j$, we quote the characteristic scales $\tilde{\beta}_j$ (in arcsec) captured by that scale, which is $2^{j}$ times the pixel size. Among the different versions of the model, the reduced $\chi^2$ values are very similar, the smallest value being achieved by the highest resolution model\footnote{Seemingly small differences in $\chi^2$ can still translate to potentially large differences in terms of Bayesian evidence. However, it is outside the scope of this work to use nested sampling to compute the Bayesian evidence, as it is too computationally expensive on a personal computer. Nevertheless, we note that nested sampling also benefits from a gradient-informed exploration of the parameter space \citep[see e.g.,][]{Albert2020jaxns}.}. An interesting aspect is visible on the finest starlet scales: for a pixel scale factor below 3, spurious isolated features appear in the solution, which is a consequence of poor regularization at small spatial scales. It is possible to address this issue by locally increasing the regularization strengths (or, alternatively, by modifying the regularization weights). Nevertheless, we find that the results of this work do not strongly depend on pixel size, and we leave such improvements for future versions of the method. Along these lines, we will investigate fully differentiable methods to optimize regularization strengths in the future.

\section{Differentiation of the $\ell_1$-norm terms \label{app:sec:l1_diff}}

The $\ell_1$-norm used to regularize \sourcepix and \lenspotpix is not differentiable when its argument is strictly zero. However, \citet{Lee2020} validates the use of the many nondifferentiable functions used in machine learning (e.g., activation functions such as ReLU), by observing that they are part of a special class of functions that allow their partial derivatives to be computed. Such functions are said to have piecewise analyticity under analytic partition.

The absolute value function at the basis of the $\ell_1$-norm belongs to this class of functions. Moreover, in practice, the function is never evaluated exactly at zero: the source or lens potential pixel values are never all exactly zero due to, for example, the presence of noise in the data. In addition, we never initialize pixel values to exactly zero on order to avoid undefined gradients at the start of the optimization.

\section{Posterior distributions for analytical profile parameters \label{app:sec:all_smooth_parameters}}

\begin{figure*}[!tbh]
    \centering
    \includegraphics[width=\linewidth]{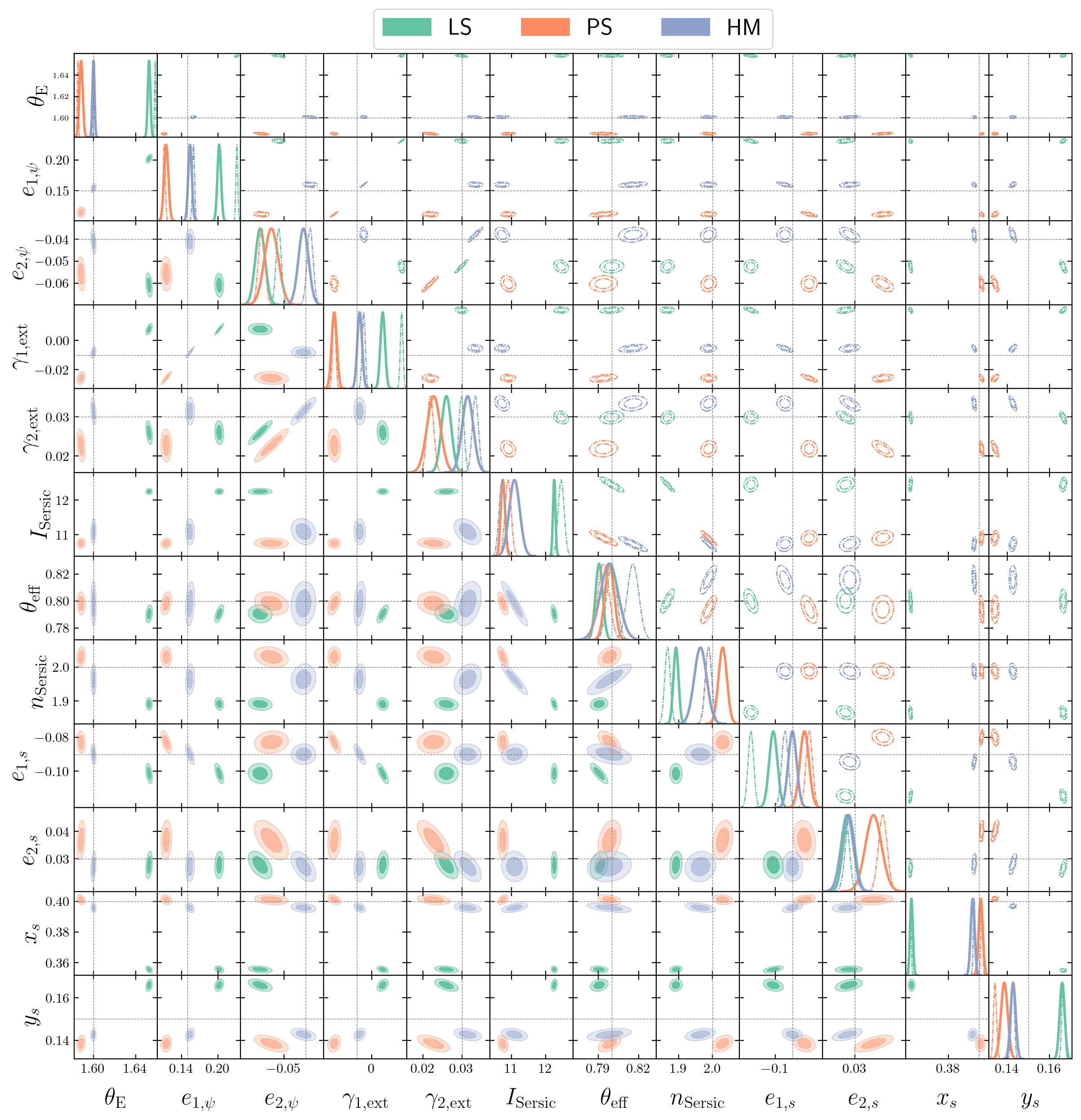}
    \caption{Posterior distributions for some analytical model parameters (smooth potential and smooth source), based on parameter covariance matrices computed either from the FIM or via HMC sampling (see Sect.~\ref{ssec:fisher_uncertainties}). The three colors correspond to the three mocks of Fig.~\ref{fig:data_summary}: localized subhalo (LS), popluation of subhalos (PS) and high-order moments (HM). Dot-dashed distributions in the upper right triangle correspond to fully smooth models, while distributions in the lower left triangle correspond to models including a pixelated component to account for perturbations of the smooth potential. Parameters are, from left to right: Einstein radius of the lens $\theta_{\rm E}$, ellipticity of the smooth lens potential $\{e_{1,\psi},e_{2,\psi}\}$, external shear components $\{\gamma_{\rm 1,ext},\gamma_{\rm 2,ext}\}$, central intensity of the source $I_{\rm Sersic}$, half-light radius $r_{\rm eff}$, Sérsic index of the source $n_{\rm Sersic}$, ellipticity of the source $\{e_{1,s},e_{2,s}\}$, position of the source $\{x_s,y_s\}$. Overall, parameter biases are reduced after including a pixelated component in the potential to model deviations from smoothness; however, some still remain, in particular for lens LS, for which turning to an analytical model informed by the pixelated model is warranted.}
    \label{app:fig:corner_plot_smooth_params}
\end{figure*}

In Sect.~\ref{ssec:smooth_parameters}, we discussed the MAP parameters of the smooth lens potential. Here we complete this discussion by showing in Fig.~\ref{app:fig:corner_plot_smooth_params} the full (smooth) parameter space, including the source parameters (Sérsic profile). For fully smooth models (upper triangle of the figure), the covariance matrix is estimated via the FIM (Sect.~\ref{ssec:fisher_uncertainties}). For the perturbed models, HMC sampling was performed to obtain the full posterior distribution. For readability, we only plot ellipses corresponding to a multivariate normal distribution based on the parameter covariance matrices.

Overall, we observe two main features from these posterior distributions. First, fully smooth models of lenses LS and PS display strong biases on almost all parameters. This is expected as the model is insufficient to capture the complexity of the lens potential. Second, for many of the parameters, biases are reduced by including the pixelated potential in the model, although the reduction is mainly due to the larger error bars. As discussed in Sect.~\ref{ssec:smooth_parameters}, these full models do not recover all smooth properties of the lens mass and source light distributions within uncertainties, despite the model fitting the data down to the noise and recovering the correct features in the pixelated potential component.

We interpret these biases as due to absorption of model residuals into the pixelated component \lenspotpix, beyond the simple deviations to the smooth potential. This hypothesis is supported by our refined models, in which we replace the pixelated component by an analytical prescription, that show no remaining bias in smooth model parameters (see e.g., Fig.~\ref{fig:smooth_params}). We expect that changing to a regularization strategy that is motivated by the specific type of perturbations is expected to mitigate these biases. For instance, based on the assumption of GRF perturbations as considered in this work, we expect the pixelated reconstruction method presented in \citet{Vernardos2022} to be less impacted by a slightly inaccurate absorption of model residuals. In a future work, we plan to explore in more detail the effect of different regularization methods on model parameters.

\end{appendix}

\typeout{get arXiv to do 4 passes: Label(s) may have changed. Rerun}

\end{document}